\documentclass[%
 aip,
 amsmath,amssymb,
preprint,%
]{revtex4-1}

\usepackage{graphicx}
\usepackage{bm}

\usepackage[utf8]{inputenc}
\usepackage[T1]{fontenc}
\usepackage{mathptmx}
\usepackage{etoolbox}
\usepackage{subfigure}
\usepackage{xcolor, soul}
\usepackage{amsmath}
\usepackage{amssymb}    

\newcommand{\csgn}{\text{csgn}}

\makeatletter
\def\@email#1#2{%
 \endgroup
 \patchcmd{\titleblock@produce}
  {\frontmatter@RRAPformat}
  {\frontmatter@RRAPformat{\produce@RRAP{*#1\href{mailto:#2}{#2}}}\frontmatter@RRAPformat}
  {}{}
}%
\makeatother
\begin{document}


\title{An investigation into the combined effect of steady and pulsatile flow within a wavy channel filled with an anisotropic porous medium}
\author{Sanchita Pramanik 
}
\author{Timir Karmakar 
}
\email{tkarmakar@nitm.ac.in}
\affiliation{ 
Department of Mathematics, National Institute of Technology Meghalaya, Sohra, Meghalaya 793108, India.
}%


\date{\today}

\begin{abstract}
We provide an analytical solution to examine the impact of steady and pulsatile flow in a symmetric wavy channel filled with an anisotropic porous medium. The flow inside the wavy channel with a porous substrate is governed by the Darcy-Brinkman equation. We used the perturbation approach to ascertain the solution to the problem, assuming that the ratio between the channel width and the wavelength is very small (i.e., $\delta^2 \ll 1$). We solved the problem up to $O(\delta^2)$ by assuming $\lambda^2 \delta^2 \ll 1$, where $\lambda$ is the ratio of anisotropic permeability. Our primary objective is to investigate how the anisotropic permeability ratio influences flow reversal near the crest and trough of a wavy wall under both steady flow conditions and when a pulsatile component is superimposed on the steady flow. In the case of steady flow, flow separation occurs near the crest region of the wavy wall, leading to the formation of recirculating zones where fluid becomes trapped. In contrast, for pulsatile flow as well as combined effects of steady and pulsatile flow, vortices form transiently at certain phases of the pulsation cycle. Subsequently, these vortices bulge into the mainstream, resulting in a complete backflow. We explored the approximate flow reversal zone using Darcy's approximation theory far from the wavy wall. The flow recirculation associated with steady flow is critical for mass transport and enhances convective mixing. On the other hand, flow reversal at different time instances can optimize oxygen transfer in devices such as membrane oxygenators. 
\end{abstract}


\maketitle


\section{Introduction}\label{Introduction}

Flow through corrugated channels is a key aspect of many fluid dynamics problems with broad scientific interest. The literature reveals that numerous studies have been investigated subject to various flow conditions corresponding to different physical situations. The flow of fluids over wavy surfaces plays a significant role in many scientific and engineering contexts, especially in mass and heat exchangers that incorporate corrugated surfaces. Such wavy surfaces are frequently observed in situations involving deformable and granular materials, such as the interaction of wind and waves in oceans or the creation of sand dunes on the ocean floor \cite{Wang2002forced,Esquivelzeta2015note}. There are numerous instances in the cardiovascular system where blood flow resembles that within a wavy channel. For example, the development and progression of arterial blockages, or stenosis, often replicate the flow patterns observed in wavy or undulating channels \cite{Griffith2009pulsatile,Sherwin2005three}. To begin, it is essential to comprehend how wavy geometry impacts flow patterns. The existing literature in this regard can be broadly classified into two categories: steady flows and unsteady flows. The latter often involves an oscillatory effect of velocity or a pulsatory effect of pressure, which plays a significant role in altering flow dynamics. This classification helps us analyze and understand the complex interactions between geometry and fluid dynamics. In this context, we will review the developments related to steady flow, followed by time-dependent flow. A primary focus will be on how flow separation occurs due to nonlinear geometries, as these geometries play a crucial role in mass transport \cite{Higdon1985stokes,Pozrikidis1987creeping,Wei2003flow,Karmakar2017note}. Tsangaris and Leiter \cite{Tsangaris1984laminar} developed a perturbation method to study viscous flows in a wavy-walled channel. To make the boundary conditions easier to implement, they transferred the channel walls into straight parallel lines. Furthermore, the stream function of the flow is expressed as a series expansion using low amplitude as a perturbed parameter. Pozrikidis \cite{Pozrikidis1987creeping} conducted a numerical investigation of steady pressure-driven Stokes flow in a wavy channel and demonstrated that the flow can separate and create a circulation in the widest part of the channel. This phenomenon depends on various parameters, including the channel width, wavelength, and amplitude of the wavy wall. Wang and Chen \cite{Wang2002forced} numerically investigated forced convection in a wavy-walled channel and found that corrugation of the channel improves the heat transfer. While small amplitude-to-wavelength ratios do not lead to a meaningful enhancement of the heat transfer, larger ratios result in a significant improvement \cite{Wang2002forced}. Scholle \cite{Scholle2004creeping} explored steady creeping flow over a wavy plate using complex function theory, examining how the plate's undulation affects the flow pattern and drag force. The boundary's waviness resembles a rough surface, and the effect of this roughness is supported by small aspect ratios in microfluidic devices, which are very often used in enhancing heat transfer \cite{Rosaguti2007low,Croce2005numerical,Rees1995boundary}. Recently, Tavakol et al. \cite{Tavakol2017extended} investigated the pressure drop and flow characteristics in channels with variable geometry using extended lubrication theory. Their study demonstrated that extended lubrication theory is a robust and accurate tool for estimating pressure drop in channels with wavy geometries.

Given its significance in multiple engineering and biological contexts, unsteady flow is also studied in relation to various physical systems. In the biomedical field, pulsatile flow has been extensively studied to better understand how blood flow patterns are affected by factors such as vascular constriction or stenosis and to assess the clinical potential of high-frequency ventilation systems used during surgery and in critical care, modeling blood oxygenators, etc. \cite{Womersley1955method,Ojha1989pulsatile,Griffith2009pulsatile,Sobey1980flow,Bandyopadhyay2012study,Charya1978pulsatile,Chaturani1986pulsatile,Chow1973laminar,Bellhouse1973high,Sobey1983occurrence,Stephanoff1980flow,Mehrotra1985pulsatile}. Heat and mass transfer in corrugated channels with pulsatile fluid flow has garnered significant interest from researchers due to its significant role in enhancing these processes. It is widely recognized that pulsation of fluid flow in a straight channel has a minimal impact on heat and mass transfer compared to that in a wavy patterned channel \cite{Nishimura1995mass,Lee1999chaotic}.

The study of steady and pulsatile flow, as well as their combination, within a wavy channel filled with a porous medium, commonly known as a packed channel, has received significant attention across various fields. These applications include biology, heat transfer enhancement, geothermal reservoirs, carbon capture and storage (CCS), the automotive industry, porous membrane oxygenators, and urine transport from the kidneys to the bladder, formation of sedimentary ripples in river channels, and dunes in deserts, among others \cite{Bellhouse1973high,Sobey1980flow,Stephanoff1980flow,Sobey1983occurrence,Guo1997pulsating,Kuznetsov2006forced,Abd2022numerical,Pozrikidis1987study,Gray2013darcy,Prasada1985mhd,Elshehawey2003effect}. Heart-lung machines are some of the most vital life support devices currently available. In the process of cardiopulmonary bypass, an oxygenator plays a critical role in delivering oxygen to the blood and eliminating carbon dioxide. The most basic form of oxygenator is the bubble oxygenator, which functions by introducing oxygen into the blood through bubbles. These devices have limited perfusion times due to blood damage resulting from direct contact with oxygen. For longer-duration bypass procedures, a membrane oxygenator is used. In this system, very often, a porous packing is used to separate the blood from the oxygen, minimizing direct exposure to the blood \cite{Sobey1980flow}. In the case of clear flow through a corrugated channel under pulsatile conditions, flow separation primarily occurs due to high Reynolds numbers or unsteady pulsation parameters such as the Womersley number or Strouhal number. However, when considering flow through a porous medium, flow separation can still occur, but an additional factor comes into play, namely, the permeability of the porous medium, which significantly influences the onset and characteristics of flow separation in such systems \cite{Wei2003flow,Feng2000lubrication,Karmakar2017note}. Here we would like to mention a few pieces of literature involving wavy corrugated geometries in the context of porous media flows. Ng and Wang \cite{Ng2010darcy} investigated the Darcy-Brinkman flow within a corrugated channel, considering the small corrugation amplitude as the perturbed parameter. Their findings indicate that the influence of the corrugation is more pronounced when the flow approaches the Stokes flow limit rather than the Darcy flow limit. Inspired by the concept that a wavy channel filled with a porous medium resembles an idealized packed fracture, which could serve as a potential pathway for carbon dioxide leakage from a geological sequestration site, Gray et al. \cite{Gray2013darcy} investigated flow inside a wavy channel filled with a Darcy porous medium. Motivated by the fact that wavy-walled channels mimic a rough surface, which may significantly enhance the heat transfer process, Chen et al. \cite{Chen2007free} studied free convection in a wavy porous cavity. More pieces of literature in this regard can be found in the study of Rao and Sivaprasad \cite{Prasada1985mhd}, Elshehawey et al. \cite{Elshehawey2003effect}, Okechi and Asghar \cite{Okechi2020darcy}, Kitanidis and Dykaar \cite{Kitanidis1997stokes}, etc. Lubrication theory, which is a special form of perturbation theory, is widely used to analyze fluid flow in narrow geometries where changes in curvature occur gradually. This approximation is particularly effective for describing the velocity and pressure fields in thin fluid films, the motion of particles within a fluid near boundaries, flow through microchannels with specified geometries, and fluid behavior over and inside surfaces with patterned topography \cite{Tavakol2017extended,Feng2000lubrication,Wei2003flow,Dewangan2025effects,Karmakar2017note,Damiano1998effect}. With the help of lubrication theory Damiano \cite{Damiano1998effect} modelled the blood flow inside capillaries, assuming the glycocalyx layer as a porous coating covering the luminal surface. As pointed out by Feng and Weinbaum \cite{Feng2000lubrication}, the gliding motion of a red cell moving over the endothelial glycocalyx is similar to that of a human skier or snowboarder skiing on compressed powder. Drawing inspiration from this analogy, they applied the concept of lubrication theory and studied the fluid flow over the glycocalyx layer, revealing that low permeability enhances the degree of flow circulation. Motivated by the physiological flows present in capillaries, Wei et al. \cite{Wei2003flow} studied the pressure-driven flow of a Newtonian fluid within a two-dimensional channel that features a wavy porous coating. Their findings revealed that when the Darcy permeability is notably low and there is a considerable difference in wall phases, the flow creates a trapped circulation eddy close to the channel's widest part. This occurrence plays an important role in the transport of solutes through the cellular boundary. Incorporating porous media into the study of pulsatile flow is another interesting topic that has drawn the attention of researchers, particularly in modeling blood vessels where fatty deposits and arterial blockages may occur. Significant contributions to this field include studies by Dawood et al. \cite{Dawood2024pulsatile,Dawood2024effect}, Sorek and Sideman \cite{Sorek1986porous}, Preziosi and Farina \cite{Preziosi2002darcy}, Ogulo and Amos \cite{Ogulu2007modeling}, etc. 

While the studies mentioned above largely focused on isotropic porous media, many real-world situations involve anisotropic porous materials where permeability differs between the horizontal and vertical directions. The anisotropy is mainly caused by the orientation and shape of the asymmetric grains forming the porous structure \cite{Rice1970anisotropic,Rees1995effect,Rees1995boundary,Bhamidipati2020shear,Karmakar2016lifting,Kohr2008green}. Numerous studies in the literature have explored the anisotropic permeability of the porous medium, particularly in the context of geothermal processes, groundwater flow through anisotropic sediments and rock formations. For example, Kvernvold and Tyvand \cite{Kvernvold1979nonlinear} studied nonlinear thermal convection, discussing heat transport and convective stability while assuming different permeabilities along the three principal axes. Bhamidipati and Woods \cite{Bhamidipati2020shear} examined the dispersion of a passive tracer in a two-dimensional pressure-driven flow through a channel, which involves anisotropic porous rock formations, and discussed the longitudinal spreading of the solute. Karmakar and Raja Sekhar \cite{Karmakar2017note} investigated flow reversal in a wavy channel filled with anisotropic porous material. They found that the anisotropic permeability of the porous medium can lead to flow separation near the crests of the wavy wall, where viscous forces are significant. Karmakar \cite{Karmakar2021} investigated unsteady Couette flow within a plane channel filled with anisotropic porous material, driven by the combined effects of a pulsating pressure gradient and the oscillatory motion of the upper plate. The study focuses on the anisotropic properties of the endothelial glycocalyx, primarily composed of bush-like collagen fibers. It analyzes the distribution of shear stress at the wall, which is significant for understanding potential damage to the arterial wall. Recently, the study of pulsatile flow within corrugated channels filled with porous media has gained significant attention from researchers, particularly in the modeling of membrane oxygenators \cite{Sobey1980flow,Stephanoff1980flow}. In such devices, porous packings are used to separate the blood from oxygen and minimize blood damage. Very often, fiber bundles are used inside membrane oxygenators, which form an anisotropic network that allows shape optimization so that efficient gas exchange occurs with minimal thrombus formation and hemolysis. Also, controlling the orientation of the fibers, the anisotropic permeability of the porous medium can allow for optimizing oxygen transfer \cite{Bhavsar2011numerical}. While previous studies on wavy-walled channels have predominantly examined flow through isotropic porous media, the impact of anisotropic properties of the porous medium, especially under pulsatile pressure conditions, remains largely unexplored. In particular, the flow separation induced by anisotropic properties with the combined effect of wall waviness and pressure pulsation presents a compelling area for investigation. This study aims to elucidate the influence of porous medium anisotropy on flow reversal within wavy channels. Gaining insight into this phenomenon could have significant implications for enhancing mass transfer efficiency and promoting convective mixing.

\section{Mathematical formulation}\label{Section 2}
\begin{figure}[h!]
    \centering
    \includegraphics[height= 2.3 in, width=4.0 in]{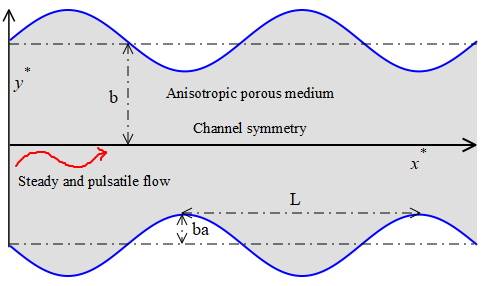}
    \caption{Schematic of physical configuration and coordinate system of the wavy-walled channel.}
    \label{Geometry}
\end{figure}

The schematic representation of the physical configuration of the problem is shown in Fig. \ref{Geometry}. We analyze a two-dimensional flow of an incompressible fluid within a wavy channel filled with an anisotropic porous medium, where the flow is driven by a periodic pressure gradient. At $y^*=0$, the channel exhibits symmetry, with the wavy bottom and upper wall positioned at $y^*=-H^{*}(x^*)$ and $y^*=H^{*}(x^*)$, respectively, where $H^{*}(x^*)=b(1+a\sin(2\pi x^*))$. We define the channel mean width as $b$, the amplitude as $ba$, and $L$ as the wavelength.


\subsection{Governing equations and boundary conditions}
We assume that the flow inside the wavy channel filled with porous medium is governed by the Darcy-Brinkman equation, supported by the equation of continuity, given respectively by
\begin{align}
    \rho_{f} \frac{\partial \textbf{V}^*}{\partial t^*}&=-\nabla {P}^* + \mu_{\textrm{eff}} \nabla^2 \textbf{V}^* - \mu \mathbf{K^{-1}} \textbf{V}^*, \\
    \nabla.\textbf{V}^*&=0,
\end{align}
where, $\textbf{V}^*=(u^*,v^*)$ is the velocity of the fluid inside the porous medium, $P^*$ is the pressure, $\mu$ is the viscosity of the fluid, $\mu_{\textrm{eff}}$ is the effective viscosity inside the porous medium, $\rho_{f}$ is the density of the fluid.

Taking into account the anisotropic nature of the porous medium, we represent the permeability $\textbf{K}$ as a second-order tensor given by
\begin{equation}
\textbf{K}=
\begin{bmatrix}
K_{x} & 0 \\
0 & K_{y} 
\end{bmatrix},
\end{equation}
where $K_{x}$ and $K_{y}$ are the permeabilities of the porous matrix in the horizontal and vertical directions, respectively.

Now, the set of governing equations will take the form,
\begin{equation}
    \rho_{f} \bigg(\frac{\partial u^*}{\partial t^*}\bigg)=-\frac{\partial P^*}{\partial x^*}+ \mu_{\textrm{eff}} \bigg(\frac{\partial^2 u^*}{\partial x^{*2}}+\frac{\partial^2 u^*}{\partial y^{*2}}\bigg)-\frac{\mu}{K_{x}} u^*,
\end{equation}
\begin{equation}
    \rho_{f} \bigg(\frac{\partial v^*}{\partial t^*}\bigg)=-\frac{\partial P^*}{\partial y^*}+ \mu_{\textrm{eff}} \bigg(\frac{\partial^2 v^*}{\partial x^{*2}}+\frac{\partial^2 v^*}{\partial y^{*2}}\bigg)-\frac{\mu}{K_{y}} v^*,
\end{equation}
\begin{equation}
    \frac{\partial u^*}{\partial x^*}+\frac{\partial v^*}{\partial y^*}=0.
\end{equation}
The fluid velocities vanish on the impermeable wall and are given by 
\begin{equation}
  \text{(i)}  \quad u^*=0, \quad v^*=0 \quad \text{at} \quad y^*=H^{*}(x^*).
\end{equation}
Considering the symmetry of the geometry, we have
\begin{equation}
  \text{(ii)}  \quad \frac{\partial u^*}{\partial y^*}=0 \quad \text{and} \quad v^*=0 \quad \text{at} \quad y^*=0.
\end{equation}
The instantaneous flow rate is given by
\begin{equation} \label{volumetric flow rate condition}
    Q^*(t^*)=\bar{Q} (1+\chi e^{i \Omega_{p} t^*})=\int^{H^{*}(x^*)}_0 u^*(x^*,y^*,t^*) dy^*,
\end{equation}
where $\bar{Q}$ is the net flow rate. 

In order to solve the above equations subject to the specified boundary conditions, we seek a solution with velocity and pressure of the form
\begin{equation}
    u^*(x^*,y^*,t^*)=u_{s}^*(x^*,y^*)+\chi \Re e (e^{i \Omega_{p} t^*} u_{p}^*(x^*,y^*)),
\end{equation}
\begin{equation}
    v^*(x^*,y^*,t^*)=v_{s}^*(x^*,y^*)+\chi \Re e (e^{i \Omega_{p} t^*} v_{p}^*(x^*,y^*)),
\end{equation}
\begin{equation}
    P^*(x^*,y^*,t^*)=P_{s}^*(x^*,y^*)+\chi \Re e (e^{i \Omega_{p} t^*} P_{p}^*(x^*,y^*)).
\end{equation}
Here, the subscripts $s$ and $p$ denote the components of steady and pulsation, respectively, $\chi$ and $\Omega_{p}$ are the amplitude and frequency of the pulsation, respectively. The notation "$\Re e$" represents the real part of a quantity.

Under the above assumptions, the corresponding governing equation for the steady part is given by
\begin{align}
    -\frac{\partial P_{s}^*}{\partial x^*}+ \mu_{\textrm{eff}} \bigg(\frac{\partial^2 u_{s}^*}{\partial x^{*2}}+\frac{\partial^2 u_{s}^*}{\partial y^{*2}}\bigg)-\frac{\mu}{K_{x}} u_{s}^*&=0, \\
     -\frac{\partial P_{s}^*}{\partial y^*}+ \mu_{\textrm{eff}} \bigg(\frac{\partial^2 v_{s}^*}{\partial x^{*2}}+\frac{\partial^2 v_{s}^*}{\partial y^{*2}}\bigg)-\frac{\mu}{K_{y}} v_{s}^*&=0, \\
    \frac{\partial u_{s}^*}{\partial x^*}+\frac{\partial v_{s}^*}{\partial y^*}&=0,
\end{align}
and for the unsteady part is given by 
\begin{align}
 -\frac{\partial P_{p}^*}{\partial x^*}+ \mu_{\textrm{eff}} \bigg(\frac{\partial^2 u_{p}^*}{\partial x^{*2}}+\frac{\partial^2 u_{p}^*}{\partial y^{*2}}\bigg)-\frac{\mu}{K_{x}} u_{p}^*&=i   \rho_{f} \Omega_{p} u_{p}^*, \\
 -\frac{\partial P_{p}^*}{\partial y^*}+ \mu_{\textrm{eff}} \bigg(\frac{\partial^2 v_{p}^*}{\partial x^{*2}}+\frac{\partial^2 v_{p}^*}{\partial y^{*2}}\bigg)-\frac{\mu}{K_{y}} v_{p}^*&=i   \rho_{f} \Omega_{p} v_{p}^*, \\
    \frac{\partial u_{p}^*}{\partial x^*}+\frac{\partial v_{p}^*}{\partial y^*}&=0.
\end{align}

We introduce the following non-dimensional variables: 
\begin{eqnarray}
 x=\frac{x^*}{L}, ~   y=\frac{y^*}{b}, ~ (u_{s},u_{p})=\frac{(u_{s}^*,u_{p}^*)}{\bar{Q}/b}, ~(v_{s},v_{p})=\frac{(v_{s}^*,v_{p}^*)}{\bar{Q}/L}, ~P=\frac{(P_{s}^*,P_{p}^*)}{{\mu \bar{Q} L}/{K_{x} b}}, ~\delta=\frac{b}{L},\\ ~Da=\frac{K_{x}}{b^2},~\lambda^2=\frac{K_{x}}{K_{y}}, ~M=\frac{\mu_{\textrm{eff}}}{\mu}, ~\textrm{Wo}=\frac{\rho_{f} \Omega_{p} b^2}{\mu}.   
\end{eqnarray}

Here, $\delta$ denotes the aspect ratio of the channel, $\lambda^2$ denotes the ratio of horizontal and vertical permeability, which represents the anisotropic permeability ratio, $Da$ is the Darcy number, $M$ is the viscosity ratio, and $\textrm{Wo}$ is the Womersley number.

After non-dimensionalization, we get that the steady part of the velocity component satisfies
\begin{align}
    M\left(\delta^2 \frac{\partial^2 u_{s}}{\partial x^2}+\frac{\partial^2 u_{s}}{\partial y^2}\right)-\alpha^2 u_{s}&=\alpha^2 \frac{\partial P_{s}}{\partial x},\\
    M\left(\delta^4 \frac{\partial^2 v_{s}}{\partial x^2}+\delta^2\frac{\partial^2 v_{s}}{\partial y^2}\right)-\alpha^2 \lambda^2 \delta^2 v_{s}&=\alpha^2 \frac{\partial P_{s}}{\partial y},\\
    \frac{\partial u_{s}}{\partial x}+\frac{\partial v_{s}}{\partial y}&=0,
\end{align}
and the corresponding pulsatile part satisfies
\begin{align} 
   M \bigg(\delta^2 \frac{\partial^2 u_{p}}{\partial x^2}+\frac{\partial^2 u_{p}}{\partial y^2}\bigg)-\xi^2 u_{p}&= \alpha^2 \frac{\partial P_{p}}{\partial x}, \label{x component pulsatile eqn} \\
   M \bigg(\delta^4 \frac{\partial^2 v_{p}}{\partial x^2}+\delta^2\frac{\partial^2 v_{p}}{\partial y^2}\bigg)-(\alpha^2 \lambda^2+\eta^2) \delta^2 v_{p}&= \alpha^2 \frac{\partial P_{p}}{\partial y}, \label{y component pulsatile eqn} \\
   \frac{\partial u_{p}}{\partial x}+\frac{\partial v_{p}}{\partial y}&=0, \label{pulsatile eqn of continuity}
\end{align}
where, $\xi^2=(\eta^2+\alpha^2)$, $\eta=\sqrt{i \hspace{0.1 cm} \textrm{Wo}}$, \quad and \quad $\alpha^2=1/Da$ .

After non-dimensionalization, we get the boundary conditions as 
\begin{equation} \label{1st BC}
\text{(i)} \quad u_{(s,p)}=0 \quad \text{and} \quad v_{(s,p)}=0 \quad \text{at} \quad y=H(x),
\end{equation}
\begin{equation} \label{2nd BC}
    \text{(ii)} \quad \frac{\partial u_{(s,p)}}{\partial y}=0 \quad \text{and} \quad v_{(s,p)}=0 \quad \text{at} \quad y=0.
\end{equation}

The non-dimensional volumetric flow rate is given by
\begin{equation} \label{Pulsatile volumetric condition}
    \int^{H(x)}_{0} u_{(s,p)}(x,y) dy=1.
\end{equation}


\section{Lubrication approximation}
For solving the boundary value problem involving a small parameter $\delta$, we employ the regular perturbation method. 
We can get the approximate solution of the problem by using the perturbation expansion in Eqs. (\ref{x component pulsatile eqn}-\ref{pulsatile eqn of continuity}). We assume that the aspect ratio of the channel is small, i.e., $\delta^2 \ll 1$, which allows us to apply the perturbation theory. Accordingly, with respect to the small perturbed parameter $\delta^{2}$, we use the following expansions of velocity and pressure 
\begin{equation}
    u_{(s,p)}=u_{(s_{0},p_{0})}+\delta^2 u_{(s_{1},p_{1})}+O(\delta^4),
\end{equation}
\begin{equation}
    v_{(s,p)}=v_{(s_{0},p_{0})}+\delta^2 v_{(s_{1},p_{1})}+O(\delta^4),
\end{equation}
\begin{equation}
    P_{(s,p)}=P_{(s_{0},p_{0})}+\delta^2 P_{(s_{1},p_{1})}+O(\delta^4).
\end{equation}
The first-order corrections appear $O(\delta^{2})$ because there are no terms of order $\delta$ in the governing equations or the boundary conditions. To solve the velocity field, we collect similar powers of $\delta^{2}$. The solution is computed up to $O(\delta^{2})$. We will present the solution for each order only for the pulsatile case, and the solution for the corresponding steady case can be obtained by setting the pulsatile parameter $\eta=0$.

\subsection{The leading-order problem}
For the pulsatile part, the leading-order governing equations (\ref{x component pulsatile eqn})-(\ref{pulsatile eqn of continuity}) become
\begin{align}
    \xi^2 u_{p_{0}}-M \frac{\partial^2 u_{p_{0}}}{\partial y^2}&=-\alpha^2 \frac{\partial P_{p_{0}}}{\partial x}, \label{Leading up0 1st eqn}\\
    \frac{\partial P_{p_{0}}}{\partial y}&=0, \label{Leading up0 2nd eqn}\\
    \frac{\partial u_{p_{0}}}{\partial x}+\frac{\partial v_{p_{0}}}{\partial y}&=0. \label{Leading up0 3rd eqn}
\end{align}

We assume that $\textrm{Wo}/\alpha^{2}\sim O(1)$. 
The leading-order boundary conditions are
\begin{equation} \label{1st BC leading Pulsatile}
\text{(i)} \quad u_{p_{0}}=0 \quad \text{and} \quad v_{p_{0}}=0 \quad \text{at} \quad y=H(x),
\end{equation}
\begin{equation} \label{2nd BC leading Pulsatile}
    \text{(ii)} \quad \frac{\partial u_{p_{0}}}{\partial y}=0 \quad \text{and} \quad v_{ p_{0}}=0 \quad \text{at} \quad y=0.
\end{equation}

The leading-order volumetric flow rate condition is
\begin{equation} \label{Leading order volumetric condition pulsatile}
    \int^{H(x)}_{0} u_{ p_{0}}(x,y) dy=1.
\end{equation}

The solution of Eq. (\ref{Leading up0 1st eqn}) subject to the above boundary conditions is given by
\begin{equation}
    u_{p_{0}}(x,y)=- \frac{\alpha^2}{\xi^2} P_{1}(x)+a_{1}(x) \cosh\bigg({\frac{\xi y}{\sqrt{M}}}\bigg)+a_{2}(x) \sinh\bigg({\frac{\xi y}{\sqrt{M}}}\bigg),
\end{equation}
\begin{equation}
  a_{1}(x)=\frac{\alpha^2}{\xi^2} \frac{P_{1}(x)}{\cosh\big({\frac{\xi H(x)}{\sqrt{M}}}\big)},  
\end{equation}
\begin{equation}
    a_{2}(x)=0,
\end{equation}

where, $P_{1}(x)=\frac{\partial P_{p_{0}}}{\partial x}$.

The constant volumetric flux condition given in Eq. (\ref{Leading order volumetric condition pulsatile}) evaluates the pressure gradient $P_{1}(x)$ as
\begin{equation}
    P_{1}(x)=\frac{\xi^2}{\alpha^2} \frac{\xi \cosh{\big(\frac{\xi H(x)}{\sqrt{M}}\big)}}{\sqrt{M} \sinh{\big(\frac{\xi H(x)}{\sqrt{M}}\big)}-\xi H(x) \cosh{\big(\frac{\xi H(x)}{\sqrt{M}}\big)}}.
\end{equation}

\subsection {The $O$($\delta^2$) problem:}
The governing equations corresponding to the first order are given by
\begin{align}
     M \bigg( \frac{\partial^2 u_{p_{0}}}{\partial x^2}+\frac{\partial^2 u_{p_{1}}}{\partial y^2}\bigg)-\xi^2 u_{p_{1}}&=\alpha^2 \frac{\partial P_{p_1}}{\partial x}, \label{First order pulsatile eqn 1} \\
     M \frac{\partial^2 v_{p_{0}}}{\partial y^2}-(\alpha^2 \lambda^2+\eta^2) v_{p_{0}}&=\alpha^2 \frac{\partial P_{p_{1}}}{\partial y}, \label{First order pulsatile eqn 2} \\
    \frac{\partial u_{p_{1}}}{\partial x}+\frac{\partial v_{p_{1}}}{\partial y} &=0. \label{First order continuity eqn}
\end{align}

The boundary conditions are
\begin{equation} \label{1st BC first order Pulsatile}
\text{(i)} \quad u_{p_{1}}=0 \quad \text{and} \quad v_{p_{1}}=0 \quad \text{at} \quad y=H(x),
\end{equation}
\begin{equation} \label{2nd BC first order Pulsatile}
    \text{(ii)} \quad \frac{\partial u_{p_{1}}}{\partial y}=0 \quad \text{and} \quad v_{p_{1}}=0 \quad \text{at} \quad y=0.
\end{equation}

The volumetric flow rate condition is
\begin{equation} \label{First order volumetric condition pulsatile}
    \int^{H(x)}_{0} u_{p_{1}}(x,y) dy=0.
\end{equation}

Eliminating pressure gradient term from Eq. (\ref{First order pulsatile eqn 1}) and (\ref{First order pulsatile eqn 2}) we get,
\begin{equation} \label{Eqn after pressure elimination}
     \frac{\partial}{\partial y} \bigg(M \frac{\partial^2 u_{p_{1}}}{\partial y^2}-\xi^2 u_{p_{1}}\bigg)=2M \frac{\partial}{\partial x}\bigg(\frac{\partial^2 v_{p_0}}{\partial y^2}\bigg)-\big(\alpha^2 \lambda^2+\eta^2\big) \frac{\partial v_{p_0}}{\partial x}.
\end{equation}

Now, using Eq. (\ref{First order continuity eqn}) we get the value of $v_{p_0}(x,y)$ as
\begin{equation} \label{expression for vp0}
    v_{p_0}(x,y)=\frac{\alpha^2}{\xi^2} \frac{d P_{1}}{dx} y-\frac{\sqrt{M}}{\xi} \frac{d a_{1}}{dx} \sinh\bigg({\frac{\xi y}{\sqrt{M}}}\bigg)+b_{1}(x).
\end{equation}

Using the boundary condition for $v_{p_0}$, we get the value of $b_{1}(x)$ as
\begin{equation}
    b_{1}(x)=0.
\end{equation}

Now, using the value of $v_{p_0}$ in Eq. (\ref{Eqn after pressure elimination}) we get

\begin{equation} \label{Eqn to solve up1}
    \frac{\partial}{\partial y} \bigg(M \frac{\partial^2 u_{p_{1}}}{\partial y^2}-\xi^2 u_{p_{1}}\bigg)=\frac{\sqrt{M}}{\xi} \bigg(\alpha^2 (\lambda^2-2)-\eta^2\bigg) \frac{d^2 a_{1}}{d x^2} \sinh\bigg({\frac{\xi y}{\sqrt{M}}}\bigg)- (\alpha^2 \lambda^2+\eta^2) \frac{\alpha^2}{\xi^2}\frac{d^2 P_{1}}{d x^2} y.
\end{equation}

Solving Eq. (\ref{Eqn to solve up1}) we get
\begin{equation} \label{solution of up1}
    u_{p_1}(x,y)=c_{1}(x) \cosh{\bigg(\frac{\xi y}{\sqrt{M}}\bigg)}+c_{2}(x) \sinh{\bigg(\frac{\xi y}{\sqrt{M}}\bigg)}+c_{3}(x)+F(x,y),
\end{equation}

where
\begin{equation}
    F(x,y)=\frac{\sqrt{M}}{2 \xi^3} \big(\alpha^2 (\lambda^2-2)-\eta^2\big) \frac{d^2 a_{1}}{d x^2} y \sinh{\bigg(\frac{\xi y}{\sqrt{M}}\bigg)}+(\alpha^2 \lambda^2+\eta^2) \bigg(\frac{\alpha^2}{2 \xi^4} \frac{d^2 P_{1}}{d x^2} y^2+\frac{y}{\xi^2} \frac{d b_{1}}{dx}\bigg).
\end{equation}

Using the boundary conditions presented in Eq. (\ref{1st BC first order Pulsatile}-\ref{First order volumetric condition pulsatile}) we get the values of $c_{1}(x)$, $c_{2}(x)$, $c_{3}(x)$ as
\begin{align}
    c_{1}(x)&=\frac{G(x)-H(x) F(x, H(x))}{H(x) \cosh{\bigg(\frac{\xi H(x)}{\sqrt{M}}\bigg)}-\frac{\sqrt{M}}{\xi} \sinh {\bigg(\frac{\xi H(x)}{\sqrt{M}}\bigg)}},\\
    c_{2}(x)&=0,\\
    c_{3}(x)&=-c_{1}(x) \cosh{\bigg(\frac{\xi H(x)}{\sqrt{M}}\bigg)}-F(x,H(x)),
\end{align}

where
\begin{equation}
    G(x)=\int ^{H(x)}_0 F(x,y) dy.
\end{equation}

By setting $\eta=0$ in the solutions previously derived for the pulsatile case, we obtained the solutions for the corresponding steady case.



\section{Results and Discussion} \label{Results and Discussion}

We will discuss three distinct cases: steady flow, pulsatile flow, referred to as Womersley flow, and the combined effect of steady and pulsatile flow, which we refer to as unsteady flow.

\begin{table*}[t]
    \centering
   \begin{tabular}{   p{4.0cm} p{6.0cm} p{4.0cm} }
      Shear stress component   &   Description   &  Definition \\
      \hline
       $\tau_{xy}^{s}$  & Steady state shear stress at the wall  &  $\left(\frac{\partial u^{s}}{\partial y}+ \delta^{2} \frac{\partial v^{s}}{\partial x}\right)$  \\
       $\tau_{xy}^{p}$  & Pulsatile shear stress & $ \left(\frac{\partial u^{p}}{\partial y}+ \delta^{2} \frac{\partial v^{p}}{\partial x}\right) e^{i \Omega_{p} t}$ \\
       $\tau_{xy}^{sp}$  & Steady and pulsatile shear stress & $\Re e \left(\tau_{xy}^{s}+\chi \tau_{xy}^{\textrm{p}}\right)$ \\
       $\tau_{avg}$     & Time-averaged shear stress &  $\Re e \left(\frac{1}{T_{p}} \int_{0}^{T_{p}}\left(\tau_{xy}^{s}+\chi \tau_{xy}^{p}\right)dt\right)$ \\
       \hline
    \end{tabular}
    \caption{Various shear stress components used in this study}
    \label{tab:Shear stress definition}
\end{table*}

\subsection{Steady Case}

\begin{figure}[h!]
    \centering
    \subfigure[$\lambda=1$]{\label{streamline steady lambda=1}\includegraphics[height=2.0 in, width=3.0 in]{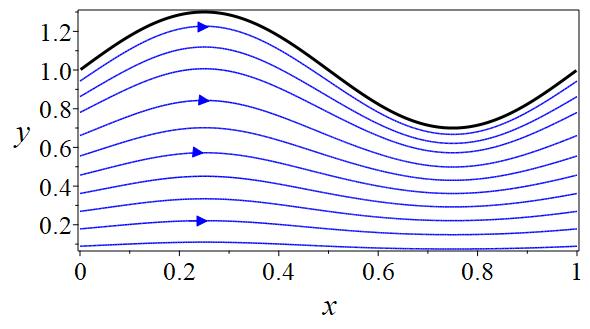}}
    \subfigure[$\lambda=1.5$]{\label{streamline steady lambda=1.5}\includegraphics[height=2.0 in, width=3.0 in]{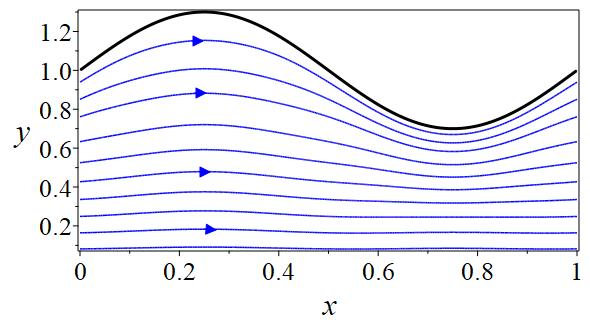}}
    \subfigure[$\lambda=1.75$]{\label{streamline steady lambda=1.75}\includegraphics[height=2.0 in, width=3.0 in]{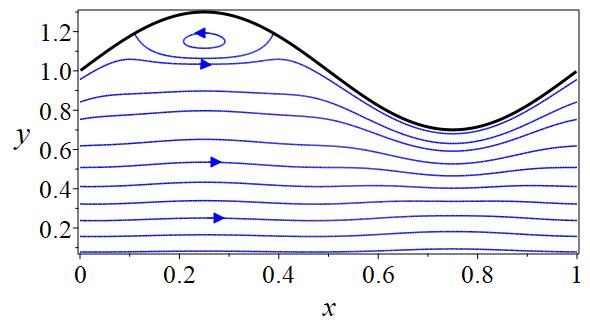}}
    \subfigure[$\lambda=2$]{\label{streamline steady lambda=2}\includegraphics[height=2.0 in, width=3.0 in]{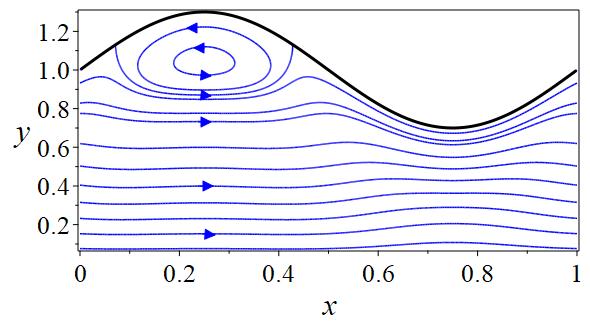}}
    \caption{Streamlines for $Da=0.01$, $\delta=0.3$, $a=0.3$, $M=1$ and various values of $\lambda$.}
    \label{Streamline Steady}
\end{figure}

Figure \ref{Streamline Steady} shows the steady state streamlines for $Da=0.01$, $\delta=0.3$, $a=0.3$, $M=1$, and various values of $\lambda$. The flow is directed along the positive $x$-axis, moving from left to right. We have illustrated the streamlines only for the upper half of the channel due to symmetry about the $x$-axis. For $\lambda=1$ and $\lambda=1.5$, the streamlines largely follow the curvature of the sinusoidal wall near the wavy boundary, while remaining nearly parallel to the $x$-axis near the centerline of the channel (see Figs. \ref{streamline steady lambda=1}-\ref{streamline steady lambda=1.5}). For larger anisotropic ratios, i.e., $\lambda=1.75$ and $\lambda=2$, the streamline separates from the main stream near the widest part of the channel (i.e., near the crest), forming a trapped recirculation zone (see Figs. \ref{streamline steady lambda=1.75}-\ref{streamline steady lambda=2}). As $\lambda$ increases, the size of the trapped recirculation region near the crest of the wavy wall also increases (see Fig. \ref{streamline steady lambda=2}).


\begin{figure}[h!]
    \centering
    \subfigure[Axial velocity at $x=0.25$]{\label{steady velocity diff lambda}\includegraphics[height=2.7 in, width=3.0 in]{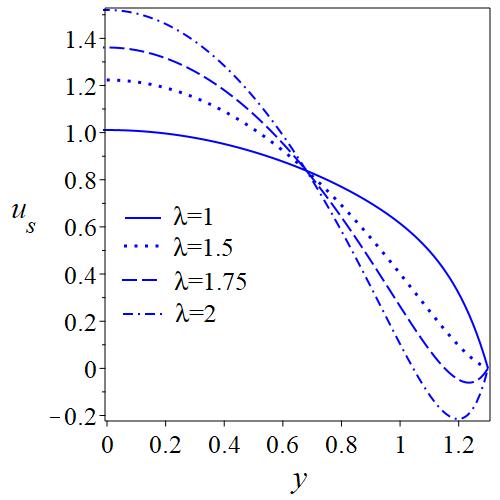}}
    \subfigure[Wall pressure gradient]{\label{pressure gradient diff lambda}\includegraphics[height=2.7 in, width=3.0 in]{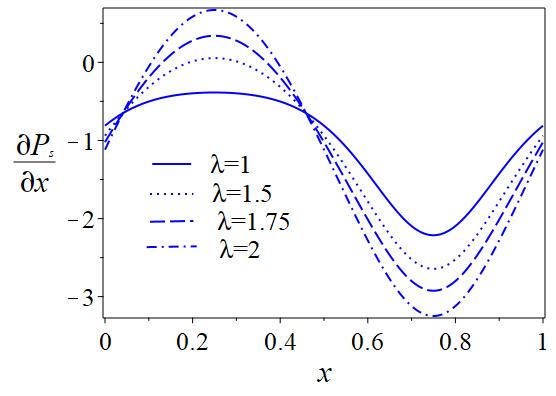}}
     \caption{Steady state axial velocity profile and wall pressure gradient for various values of $\lambda$ when the parameter values are $\delta=0.3$, $a=0.3$, $Da=0.01$, $M=1$.}
    \label{steady velocity and pressure grad}
\end{figure}

To comprehend the flow reversal phenomena in the wavy channel, it is essential to examine the velocity, the wall pressure gradient, and the shear stress. The effect of the anisotropic permeability ratio on the velocity and pressure gradient under steady state conditions is presented in Fig. \ref{steady velocity and pressure grad}.
Figure \ref{steady velocity diff lambda} depicts the velocity profile at the crest, $x=0.25$. The axial velocity is at the onset of sign change for $\lambda=1.5$, and eventually changes its sign for $\lambda=1.75$ and $2$, for which the velocity remains positive in the middle and becomes negative at the crest to maintain a constant volumetric flow rate. For a fixed value of the Darcy number ($Da$), increasing the value of $\lambda$ leads to a reduced permeability in the vertical direction. The flow encounters increased resistance, and an unfavorable (adverse) pressure gradient arises at high values of $\lambda$, resulting in flow reversal near the crest so that the overall volumetric flow rate remains constant in the cross section. 
Figure \ref{pressure gradient diff lambda} illustrates the pressure gradient at the wall of the wavy channel. The pressure gradient is negative or favorable for smaller values of the anisotropic permeability ratio ($\lambda=1,1.5$). However, for $\lambda=1.75,2$, the pressure gradient reverses its sign at the crest. An adverse pressure gradient has an opposing influence on fluid flow. As a result, the flow reversal occurs at the crest as the fluid struggles to overcome the established adverse pressure gradient.

\begin{figure}[h!]
    \centering
    \includegraphics[height= 2.6 in, width=3.0 in]{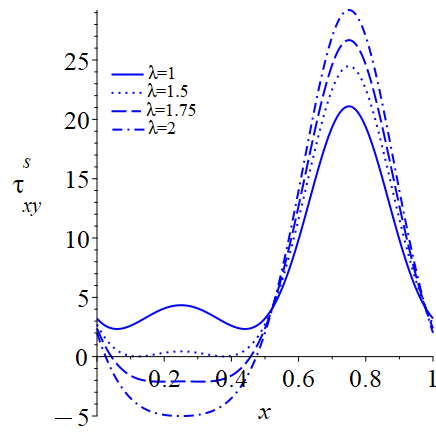}
    \caption{Steady state wall shear stress for $\delta=0.3$, $a=0.3$, $M=1$, $Da=0.01$.}
    \label{Steady shear stress}
\end{figure}

Flow separation near the crest of the wavy wall can be interpreted by analyzing the shear stress distribution across the wall. Table \ref{tab:Shear stress definition} provides the definitions of various components of shear stress used in this study. We plot the impact of the anisotropic permeability ratio ($\lambda$) on the wall shear stress in Fig. \ref{Steady shear stress} for $\delta=0.3$, $a=0.3$, $M=1$, and $Da=0.01$. The wall shear stress is higher in the constricted (convex) region and lower in the hollow (concave) region of the channel. For smaller values of $\lambda$, the wall shear stress remains positive throughout the channel. As $\lambda$ increases, the magnitude of the wall shear stress decreases, particularly near the crest beneath the concave region, due to reduced flow velocities (see Fig. \ref{steady velocity diff lambda}). For $\lambda=1.5$, the wall shear stress is on the verge of changing signs. For $\lambda=1.75$ and $2$, the wall shear stress becomes negative over a small region near $x=0.25$, indicating the occurrence of flow separation (see Fig. \ref{Steady shear stress}).


\begin{figure}[h!]
    \centering
    \includegraphics[height= 2.8 in, width=3.2 in]{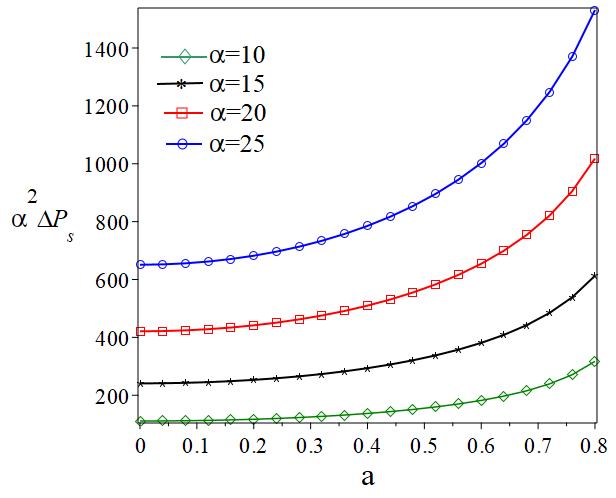}
    \caption{Steady state pressure drop at the wall for $\delta=0.3$, $a=0.3$, $M=1$.}
    \label{Steady pressure drop}
\end{figure}

The steady state pressure drop is shown in Fig. \ref{Steady pressure drop} for various values of $\alpha$. We can observe that when $\alpha$ grows, indicating a reduction in Darcy permeability, the pressure drop increases rapidly. In the current context, we define the steady state pressure drop as a multiple of $\alpha^{2}$, expressed as $\alpha^{2} \Delta P_{s}$, to capture the actual contribution for large values of $\alpha$. This behavior aligns with the findings of Wei et al. \cite{Wei2003flow}, Feng \& Weinbaum \cite{Feng2000lubrication}, who state that for large $\alpha$, the appropriate velocity scale should be $\bar{Q}/b \alpha^{2}$ in the axial direction and $\bar{Q}/L \alpha^{2}$ in the vertical direction. With these modified scaling factors, one can derive the actual pressure drop $\Delta P_{s}$. An increased value of $\alpha$, which represents reduced permeability, limits fluid movement within the channel, resulting in heightened resistance to flow and an increased pressure drop within the channel. Amplitude refers to the height of the features of the surface roughness, including peaks and valleys. An increase in amplitude indicates greater surface roughness, which enhances frictional resistance to fluid flow. As a result, this increased resistance leads to a higher pressure drop along the flow path \cite{Croce2005numerical}.


\subsection{Pulsatile Case (Womersley)}

\begin{figure}[h!]
    \centering
    \subfigure[$t=0$]{\label{streamline t=0}\includegraphics[height=1.5 in, width=2.0 in]{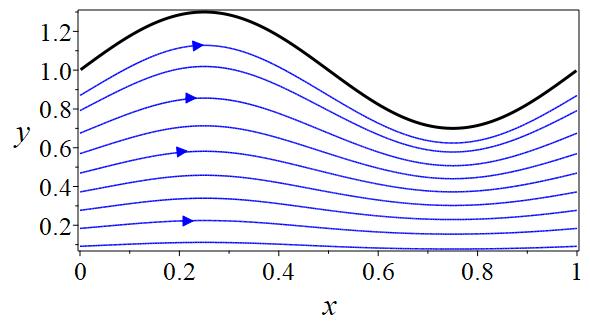}}
    \subfigure[$t=T_{p}/4$]{\label{t=T/4}\includegraphics[height=1.5 in, width=2.0 in]{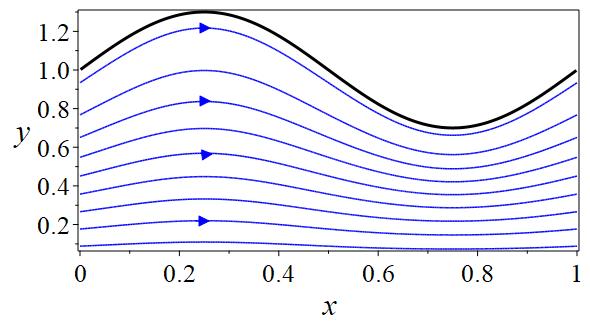}}
    \subfigure[$t=2T_{p}/4$]{\label{t=2T/4}\includegraphics[height=1.5 in, width=2.0 in]{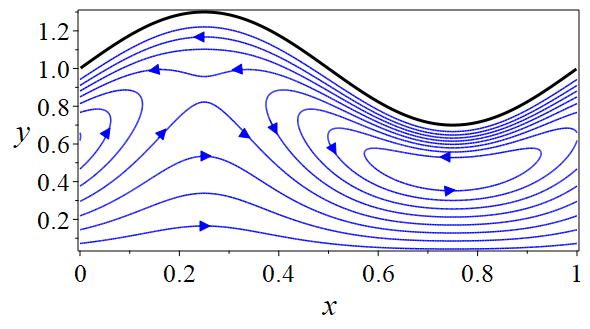}}
    \subfigure[$t=3T_{p}/4$]{\label{t=3T/4}\includegraphics[height=1.5 in, width=2.0 in]{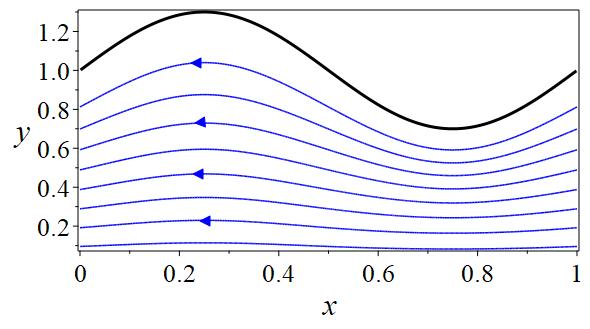}}
    \subfigure[$t=4T_{p}/4$]{\label{t=4T/4}\includegraphics[height=1.5 in, width=2.0 in]{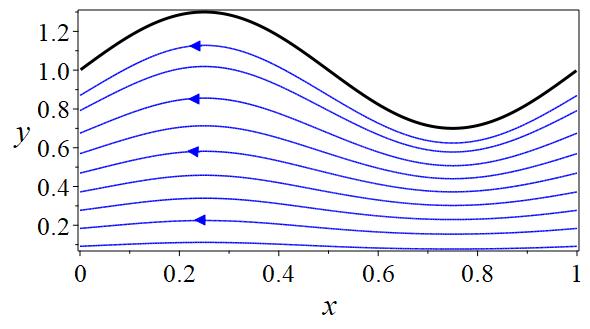}}
    \subfigure[$t=5T_{p}/4$]{\label{t=5T/4}\includegraphics[height=1.5 in, width=2.0 in]{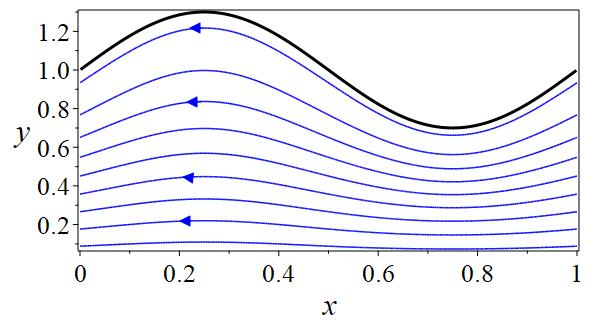}}
   \subfigure[$t=6T_{p}/4$]{\label{t=6T/4}\includegraphics[height=1.5 in, width=2.0 in]{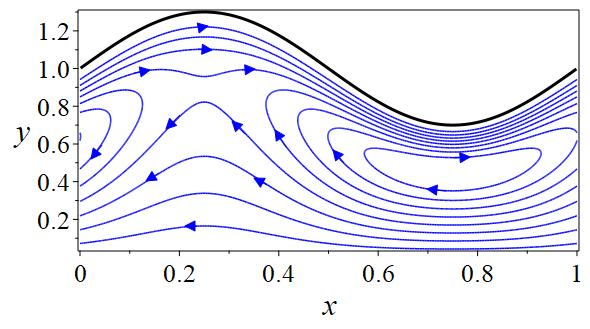}}
    \subfigure[$t=7T_{p}/4$]{\label{t=7T/4}\includegraphics[height=1.5 in, width=2.0 in]{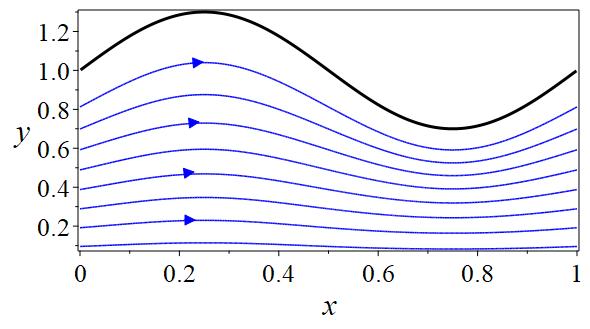}}
    \subfigure[$t=8T_{p}/4$]{\label{t=8T/4}\includegraphics[height=1.5 in, width=2.0 in]{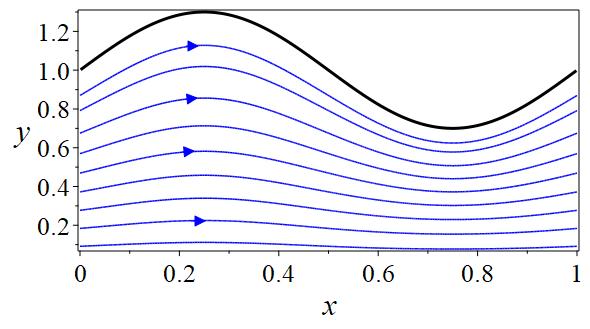}}
    \caption{Streamlines for full period of pulsation when $Da=0.01$, $\delta=0.3$, $a=0.3$, $M=1$, $\textrm{Wo}=100$ and $\lambda=1$.}
    \label{Streamline pulsation diff t}
\end{figure}

In this section, we examine the flow behavior when it is solely influenced by periodic pressure pulsations. Since flow is pressure-driven, temporal variation in the pressure gradient significantly affects the resulting flow patterns. Figure \ref{Streamline pulsation diff t} presents a sequence of computed streamlines for the parameters $Da=0.01$, $\delta=0.3$, $a=0.3$, $M=1$, $\textrm{Wo}=100$, and $\lambda=1$, illustrating how the flow evolves over time within one complete pulsation cycle. The flow exhibits a periodic back-and-forth motion, reversing direction over the course of a full pulsation cycle. At the beginning of the pulse, at $t=0$, the fluid starts from rest and flows through the channel, adhering to the shape of the wall. When the time reaches the next pulse, $t=T_{p}/4$, a weak adverse pressure gradient forms (see Figs. \ref{streamline t=0}-\ref{t=T/4}). However, this adverse pressure gradient is not strong enough to change the flow direction, so the fluid maintains its momentum and continues to move forward along the mainstream. At $t=2 T_{p}/4$, a significant eddy forms near the crest region of the channel, resulting in two co-rotating vortices beneath the apex of the hollow shifting towards the trough (i.e., $x=0.75$) (see Fig \ref{t=2T/4}). By the time $t=3T_{p}/4$ is reached, the vortices expand into the mainstream, causing a complete flow reversal from right to left throughout the channel (see Fig. \ref{t=3T/4}). In subsequent time instances, $t=4T_{p}/4, 5T_{p}/4, 6T_{p}/4, 7T_{p}/4$, similar flow behaviors are observed, mirroring those seen at $t=0, T_{p}/4, 2T_{p}/4, 3T_{p}/4$, but in the opposite direction (see Figs. \ref{streamline t=0}-\ref{t=7T/4}). At $t=2T_{p}$, the flow pattern returns to its initial state observed at $t=0$, indicating the completion of the full pulsation cycle (see Figs. \ref{streamline t=0} and \ref{t=8T/4}).

\begin{figure}[htp!]
    \centering
    \subfigure[Velocity at $x=0.75$]{\label{axial velocity fully pulsatile}\includegraphics[height=2.6 in, width=2.8 in]{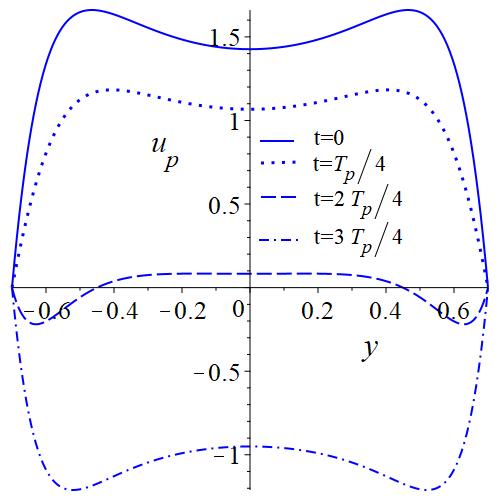}}
    \subfigure[Wall pressure gradient]{\label{axial pressure gradient fully pulsatile}\includegraphics[height=2.6 in, width=2.8 in]{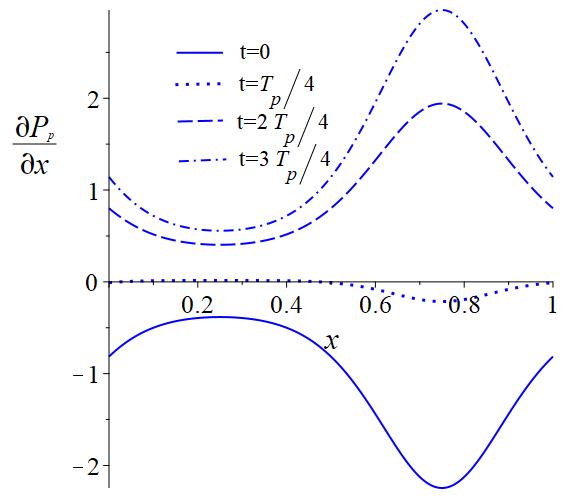}}
    \caption{Velocity and pressure gradient for the parameters $\delta=0.3$, $a=0.3$, $Da=0.01$, $M=1$, $\lambda=1$, $\textrm{Wo}=100$.}
    \label{axial velocity and pressure gradient}
\end{figure}

The flow behavior in response to flow alterations can be more easily understood in terms of the corresponding axial velocities and wall pressure gradient. Figure \ref{axial velocity and pressure gradient} presents the axial velocity and pressure gradient at four representative time instances $t=0, T_{p}/4, 2T_{p}/4$, and $3T_{p}/4$. For subsequent time instances, such as $t=4T_{p}/4, 5T_{p}/4, 6T_{p}/4, 7T_{p}/4$, the flow behavior remains unchanged except that the direction of the flow (as shown in Figs. \ref{t=4T/4}-\ref{t=7T/4}). To analyze the eddy formation, we plotted the axial velocity at the trough region ($x=0.75$) in Fig. \ref{axial velocity fully pulsatile}. At time instances $t=0$ and $T_{p}/4$, the velocity remains positive, indicating no flow reversal. However, at $t=2T_{p}/4$, the velocity changes signs and becomes negative near the boundary, indicating flow reversal. By $t=3T_{p}/4$, the velocity across the channel becomes predominantly negative, which shows a complete alteration in flow direction. This reversal signifies that the flow is now directed from right to left. The variations in axial velocity over various time intervals show how the flow evolves over time. Since flow is primarily pressure-driven, the role of axial pressure gradient becomes particularly significant, as it directly influence both the onset of flow reversal and the development of eddy structure.
Figure \ref{axial pressure gradient fully pulsatile} depicts the axial pressure gradient along the wall at the same selected time instances as that of velocity. Pulsatile flow is distinguished by forward and backward flow, in which the driving pressure changes, causing the fluid to accelerate and decelerate frequently, and, as a consequence, influencing the flow direction and sign change of the velocity profile over time. A favorable pressure gradient causes the fluid to flow at $t=0$ and $T_{p}/4$ (the beginning of the cycle), pushing the fluid forward and leading to a fully positive velocity profile. Furthermore, it is clear from Fig. \ref{axial pressure gradient fully pulsatile} that the pressure gradient is positive at $t=2T_{p}/4,3T_{p}/4$, resulting in an unfavorable (positive) pressure gradient. Since the adverse pressure gradient causes flow reversal near the trough region of the wavy channel, the velocity changes sign near the trough at $t=2T_{p}/4$ (see Fig. \ref{t=2T/4}). The velocity profile becomes negative at $t=3T_{p}/4$ due to an increase in the magnitude of the positive pressure gradient, which is strong enough to totally reverse the flow direction (see Fig. \ref{t=3T/4}).

\begin{figure}[h!]
    \centering
    \subfigure[$t=0$]{\label{streamline t=0, lambda=2}\includegraphics[height=1.9 in, width=3.0 in]{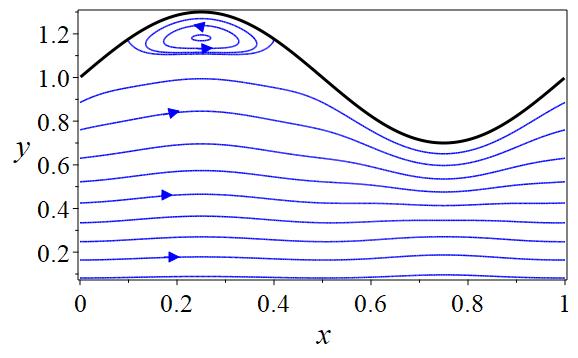}}
    \subfigure[$t=T_{p}/4$]{\label{stream t=T/4, lambda=2}\includegraphics[height=1.9 in, width=3.0 in]{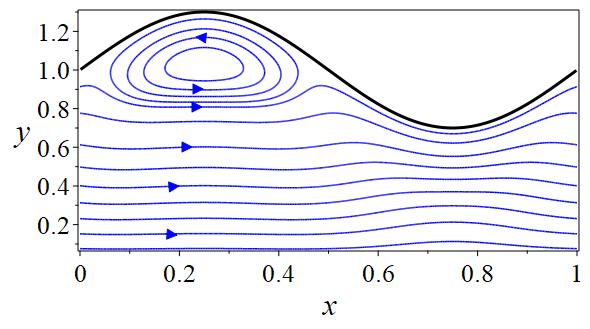}}
    \subfigure[$t=2T_{p}/4$]{\label{stream t=2T/4, lambda=2}\includegraphics[height=1.9 in, width=3.0 in]{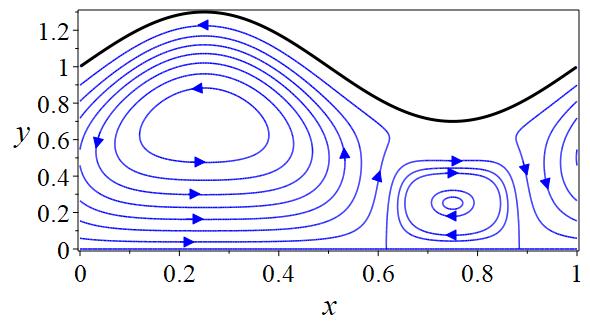}}
    \subfigure[$t=3T_{p}/4$]{\label{stream t=3T/4, lambda=2}\includegraphics[height=1.9 in, width=3.0 in]{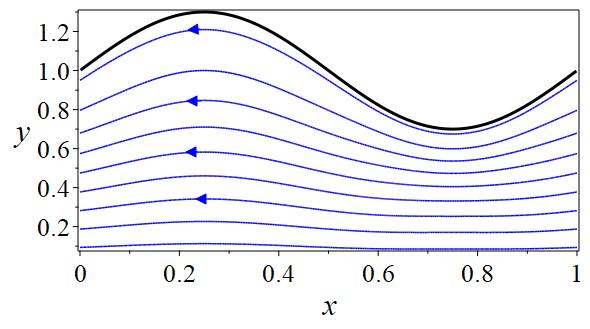}}
    \caption{Streamlines for $Da=0.01$, $\delta=0.3$, $a=0.3$, $M=1$, $\textrm{Wo}=100$, $\lambda=2$ for various time.}
    \label{Streamline only pulsatile lambda 2}
\end{figure}

To investigate the impact of anisotropic permeability on the pulsatile flow pattern, we have plotted the streamline at time instances $t=0, T_{p}/4, 2T_{p}/4$, and $3T_{p}/4$, for the case when $\lambda=2$. Unlike the classical isotropic case ($\lambda=1$), where no eddy formation is observed at time instances $t=0$ and $T_{p}/4$, the anisotropic case ($\lambda=2$) exhibits noticeable circulation near the crest region of the wavy channel (see Figs. \ref{streamline t=0, lambda=2}-\ref{stream t=T/4, lambda=2}). This indicates that a higher anisotropic permeability ratio can induce flow circulation even when no such circulation is present in the corresponding isotropic case. At $t=T_{p}/4$, the circulation zone is larger compared to that at $t=0$. By $t=2T_{p}/4$, the circulation expands to occupy nearly the entire crest region. At this stage, two counter-rotating vortices are observed: one near the crest, rotating counterclockwise, and another near the trough, rotating clockwise (see Fig. \ref{stream t=2T/4, lambda=2}). At $t=3T_{p}/4$, all developed circulation (in the previous time instances) merges into the main stream and disappears, resulting in a reversal of the flow direction, i.e., from right to left (see Fig. \ref{stream t=3T/4, lambda=2}).

\begin{figure}[h!]
    \centering
    \subfigure[Velocity at $x=0.25$]{\label{axial velocity crest lambda 2}\includegraphics[height=2.6 in, width=2.8 in]{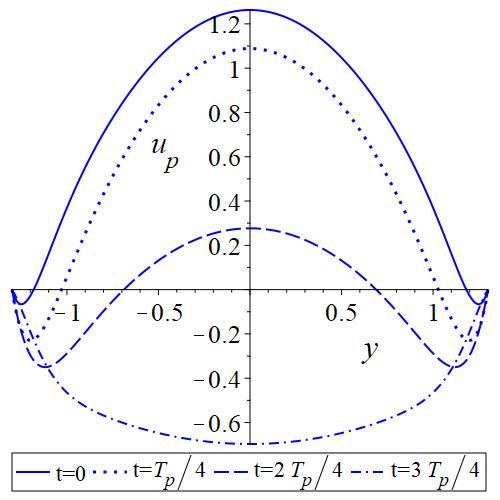}}
    \subfigure[Pressure gradient at the wall]{\label{axial pressure gradient lambda 2}\includegraphics[height=2.6 in, width=2.8 in]{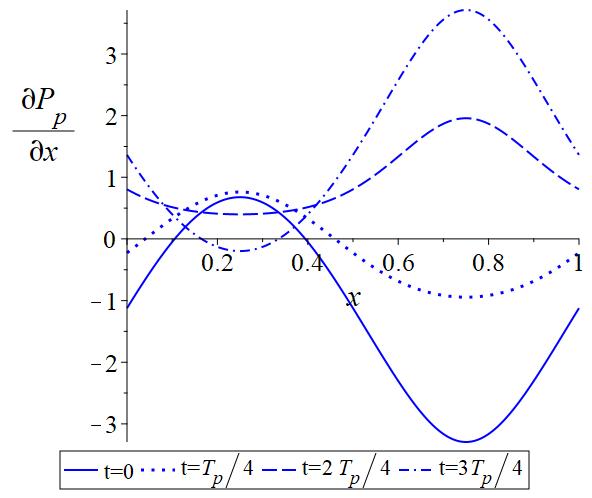}}
    \caption{Axial velocity profile and pressure gradient for different time instances when the parameter values are $\delta=0.3, a=0.3, Da=0.01, M=1, \lambda=2, \textrm{Wo}=100$.}
    \label{axial velocity crest and pressure gradient lambda 2}
\end{figure}

The underlying cause of flow circulation due to anisotropy can be better understood by examining the corresponding axial velocity and pressure gradient at various time instances. We show the axial velocity profile at the crest ($x=0.25$) in Fig. \ref{axial velocity crest lambda 2} for various time instances $t$ when the parameter values are $\delta=0.3$, $a=0.3$, $Da=0.01$, $M=1$, $\lambda=2$, and $\textrm{Wo}=100$. At the time instances $t=0$, $T_{p}/4$, and $2T_{p}/4$, the axial velocity becomes negative near the wavy boundary, where viscous effects dominate, while remaining positive near the center of the channel, where the Darcy effect is more significant (see Fig. \ref{axial velocity crest lambda 2}). As time progresses from $t=0$ to $t=2T_{p}/4$, the region of negative velocity near the boundary expands. By $t=3T_{p}/4$, a complete backflow is observed across the channel. Figure \ref{axial pressure gradient lambda 2} illustrates the variation of the pressure gradient along the wall. Within the viscous (Brinkman) layer, the shear stress induced by viscosity exerts a retarding influence on the flow. Additionally, for the anisotropic case ($\lambda=2$), the flow encounters increased resistance due to anisotropy of the porous medium. This added resistance weakens the flow's ability to overcome adverse (positive) pressure gradients, leading to flow reversal near the wavy boundary. When the magnitude of the positive pressure gradient becomes sufficiently strong, it forces the flow to reverse direction throughout the entire channel.

\subsection{Combined effect of steady and pulsation (unsteady)}

This section examines the flow behavior when a steady flow is superimposed with the pulsatile component. We refer to this combined flow condition as unsteady flow. One noticeable thing is that while the instantaneous flow rate across the channel remains constant, it varies with time in an oscillatory manner.

\begin{figure}[h!]
    \centering
    \subfigure[$t=0$]{\label{Velocity t=0}\includegraphics[height=2 in, width=2.0 in]{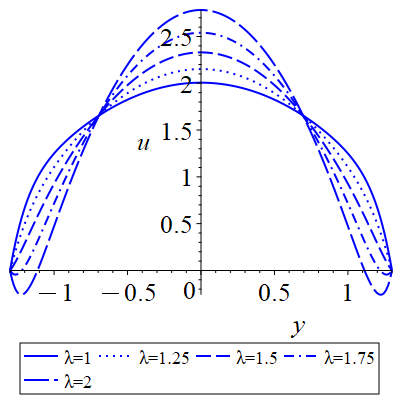}}
    \subfigure[$t=T_{p}/4$]{\label{Velocity t=T/4}\includegraphics[height=2 in, width=2.0 in]{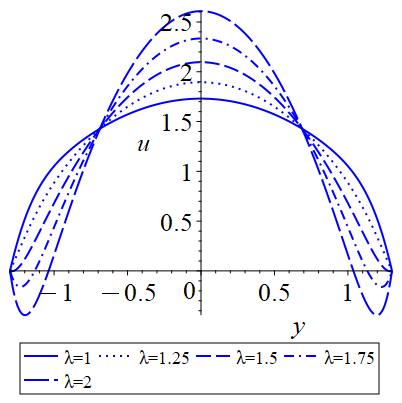}}
    \subfigure[$t=2T_{p}/4$]{\label{Velocity t=2T/4}\includegraphics[height=2 in, width=2.0 in]{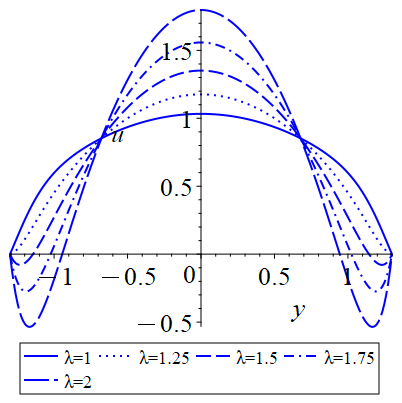}}
    \subfigure[$t=3T_{p}/4$]{\label{Velocity t=3T/4}\includegraphics[height=2 in, width=2.0 in]{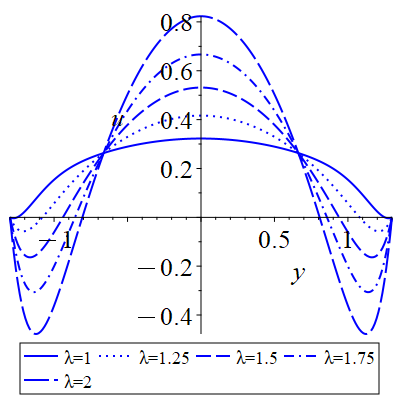}}
    \subfigure[$t=4T_{p}/4$]{\label{Velocity t=4T/4}\includegraphics[height=2 in, width=2.0 in]{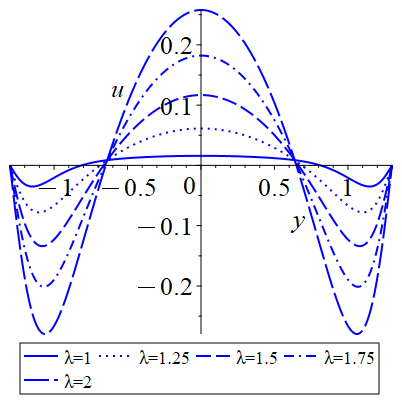}}
    \subfigure[$t=5T_{p}/4$]{\label{Velocity t=5T/4}\includegraphics[height=2 in, width=2.0 in]{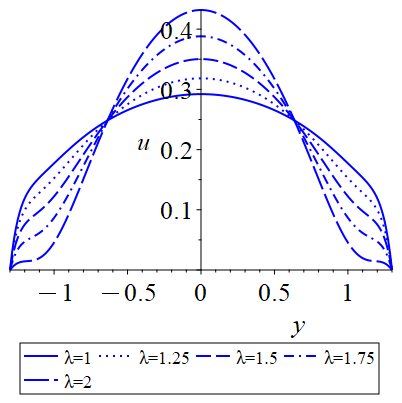}}
   \subfigure[$t=6T_{p}/4$]{\label{Velocity t=6T/4}\includegraphics[height=2 in, width=2.0 in]{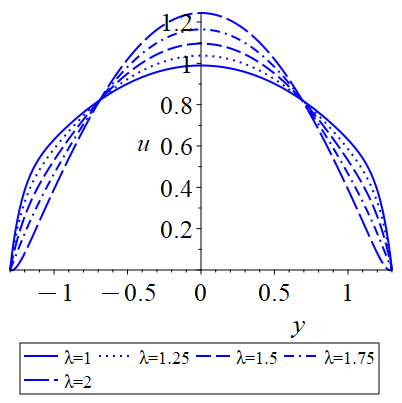}}
    \subfigure[$t=7T_{p}/4$]{\label{Velocity t=7T/4}\includegraphics[height=2 in, width=2.0 in]{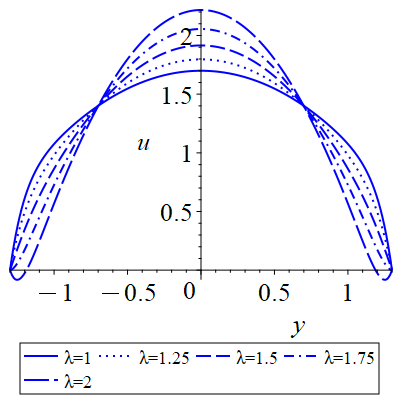}}
    \caption{Velocity at $x=0.25$ for different time $t$ and $Da=0.01$, $\delta=0.3$, $a=0.3$, $M=1$, $\textrm{Wo}=100$, $\chi=1$ and various values of $\lambda$.}
    \label{Velocity Steady and Pulsatile}
\end{figure}

Figure \ref{Velocity Steady and Pulsatile} shows a sequence of velocity profiles captured over a full pulsation cycle. Each figure depicts the velocity distribution for different anisotropic permeability ratios, highlighting how the anisotropy of the porous medium influences the flow dynamics throughout the cycle. The results indicate that the initiation of flow circulation is dependent on the influence of time and anisotropic permeability. Starting from $t=0$, we observe that the flow reversal initially occurs for higher values of the anisotropic permeability ratio. As time progresses, the flow reversal zone gradually expands, and at later time instances, small flow reversal zones also appear for lower anisotropic permeability ratios. This trend continues until the time reaches $t=4T_{p}/4$, where we observe flow reversal at the crest region for all values of the anisotropic permeability ratio. Beyond this point, the interplay between the steady component, which drives the flow forward, and the pulsatile component, which induces back-and-forth motion, leads to the disappearance of the flow reversal zones. However, they reappear near the small portion of the crest as time advances when the developed adverse pressure gradient is strong enough to reverse the weakened flow. This weakening occurs when $\lambda=2$ due to additional resistance caused by anisotropy.

\begin{figure}[h!]
    \centering
    \subfigure[Pressure gradient]{\label{Steady and pulsatile pressure grad diff t}\includegraphics[height=2.5 in, width=2.7 in]{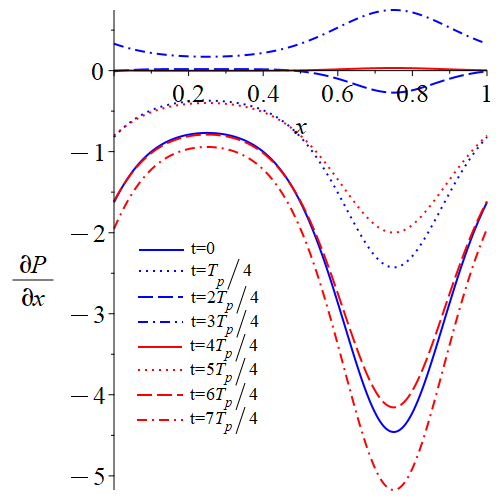}}
    \subfigure[Shear stress]{\label{Shear stress steady and pulsatile diff t}\includegraphics[height=2.5 in, width=2.7 in]{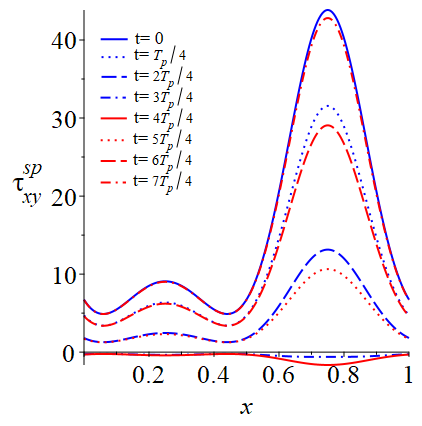}}
    \caption{Wall pressure gradient and shear stress plot for different time, $t$, for the parameters $\delta=0.3$, $Da=0.01$, $a=0.3$, $\lambda=1$, $M=1$, $\chi=1$, $\textrm{Wo}=100$.}
    \label{Steady and pulsatile pressure gradient and shear stress}
\end{figure}

Figure \ref{Steady and pulsatile pressure grad diff t} illustrates the variation in the pressure gradient across the wall in different time intervals for the parameters $\delta=0.3$, $Da=0.01$, $a=0.3$, $\lambda=1$, $M=1$, $\chi=1$, and $\textrm{Wo}=100$. The figure shows the impact of the wall pressure gradient on the flow behavior resulting from the superposition of steady and pulsatile components. It is evident that at time instances $t=0$, $T_{p}/4$, $2T_{p}/4$, $5T_{p}/4$, $6T_{p}/4$, and $7T_{p}/4$, the pressure gradient is negative. At $t=3T_{p}/4$ and $t=4T_{p}/4$, an adverse pressure gradient developed, which caused flow reversal. These results reveal that when steady and pulsatile effects are considered collectively, the timing of flow reversal and the formation of circulation differ from that observed in the purely pulsatile case.
Figure \ref{Shear stress steady and pulsatile diff t} depicts the wall shear stress produced by the interplay of steady and pulsatile fluid flow at different time intervals. Shear stress becomes negative for the values $t=3T_{p}/4$ and $t=4T_{p}/4$, while it remains positive for all other values of $t$.

\subsubsection{Pulsatile boundary layer thickness} \label{Pulsatile boundary layer thickness}

The pulsatile boundary layer is a thin layer adjacent to the wall where shear stress and viscous effects play a significant role. In a clear flow situation, the thickness of this layer is solely determined by the pulsation frequency of the flow, which is captured by the dimensionless Womersley number. This number characterizes the balance between transient inertial and viscous forces and thus governs the penetration depth of oscillatory velocity fluctuations in the fluid. In contrast, when the flow occurs within an anisotropic porous medium, the thickness of the pulsatile boundary layer is influenced by a greater number of interplay factors. In addition to the pulsatile frequency, it is significantly dependent on various properties of the porous medium, including the permeability and viscosity ratio.

From Eq. (\ref{x component pulsatile eqn}), the order of magnitude analysis gives,
\begin{equation}
    \frac{M}{y^2} \sim (\eta^2+\alpha^2),
\end{equation}
which implies,
\begin{equation}
    \frac{1}{y^2}=\frac{(i \textrm{Wo}+\alpha^2)}{M}.
\end{equation}
If we consider the pulsatile boundary layer thickness as $\delta_{\textrm{puls}}$, then we get,
\begin{equation} \label{Boundary layer thickness}
    \frac{1}{\delta_{\textrm{puls}}}=\Re e \bigg(\sqrt{\frac{i \textrm{Wo}+\alpha^2}{M}}\bigg).
\end{equation}
One may obtain the pulsatile boundary layer thickness as a function of porous media parameters as $\delta_{\textrm{puls}}=\frac{1}{S}$, where
\begin{equation} \label{Modified pulsatile boundary layer thickness}
    S=\frac{1}{\sqrt{M}} (\alpha^4+\textrm{Wo}^2)^{\frac{1}{4}} \cos{\bigg(\frac{1}{2} \tan^{-1}\bigg(\frac{\textrm{Wo}}{\alpha^2}\bigg)\bigg)}.
\end{equation}
When $\alpha \rightarrow 0$ and $M=1$, we have $\delta_{\textrm{puls}}=\sqrt{2/\textrm{Wo}}$, which is the typical oscillatory boundary layer thickness for clear flow situations.

\begin{figure}[h!]
    \centering
   \subfigure[$\textrm{Wo}=100$]{\label{Pulsatile shear y different t high Wo}\includegraphics[height=3.0 in, width=3.0 in]{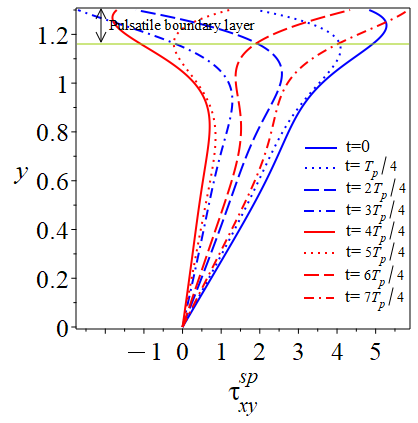}}
   \subfigure[$\textrm{Wo}=1$]{\label{Pulsatile shear y different t low Wo}\includegraphics[height=3.0 in, width=3.0 in]{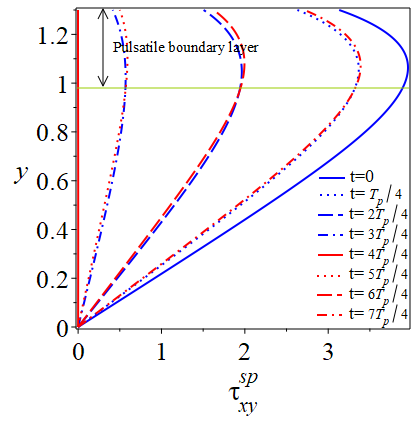}}
    \caption{Shear stress distribution along the channel width at $x=0.25$ for the full period of pulsation when the parameter values are $\delta=0.3$, $Da=0.1$, $a=0.3$, $\lambda=1$, $M=1$, $\chi=1$.}
    \label{Pulsatile high and low Wo shear stress function y}
\end{figure}

Figure \ref{Pulsatile high and low Wo shear stress function y} represents the shear stress distribution as a function of depth when the flow is driven by both steady and an oscillatory (pulsatile) pressure gradient. These figures reveal that the influence of pulsation is most significant near the wall, within a thin region commonly referred to as the pulsatile boundary layer. Within this thin layer, the shear stress distribution curve fluctuates more, highlighting the strong influence of pulsation close to the wall. These oscillations gradually diminish with depth, and the shear stress profile becomes almost linear. When the Womersley number is high ($\textrm{Wo}=100$), oscillatory effects dominate; in this case, the shear stress fluctuations are sharp and localized near the boundary inside a thin layer. In contrast, when the Womersley number is low ($\textrm{Wo}=1$), viscous effects prevail, allowing oscillations to penetrate to a larger depth inside the channel.

\section{Hydrodynamic behavior far from the wavy wall} \label{Darcy behavior far from wall}

The flow behavior close to a channel wall, where viscous forces are the most significant, has been the subject of several investigations from the past few decades. In his pioneering work, Brinkman \cite{Brinkman1949calculation} proposes a modification to Darcy's law by incorporating a viscous shear term, now known as the Brinkman term, to describe the flows in the vicinity of boundaries. Later on, Taylor \cite{Taylor1971model} and Nield \& Bejan \cite{Nield1983boundary} examined limitations of the Brinkman equation, especially in the context of the no-slip boundary condition at a solid wall. In the context of the liquid-porous interface, Hill \& Straughan \cite{Hill2008poiseuille} analyzed Poiseuille flow over a porous medium and introduced the concept of a two-layer structure. This structure consists of a Brinkman layer adjacent to the boundary, where the viscous effects are significant, and a Darcy region located farther away, where these viscous effects are negligible. Collectively, these studies suggest that in the vicinity of the wavy wall, the flow is described by the Brinkman equation, incorporating both viscous and Darcy effects. As the distance from the wall increases and viscous effects diminish, the flow asymptotically approaches the Darcy regime \cite{Karmakar2017note}. The corresponding Darcy equation can be obtained by neglecting the viscous term simply by omitting the second-order derivative term from the Brinkman equation. As a result, we have the following equation for the leading order from Eq. (\ref{Leading up0 1st eqn}):
\begin{equation}
\left(\alpha^{2}+\eta^{2}\right)u_{p_{0}}\sim -\alpha^{2}P_{1}(x),    
\end{equation}
i.e., 
\begin{equation}
u_{p_{0}}\sim -\frac{\alpha^{2}}{\alpha^{2}+\eta^{2}} P_{1}(x),    
\end{equation}
\begin{equation}
v_{p_{0}} \sim  \frac{\alpha^{2}}{\alpha^{2}+\eta^{2}} \frac{dP_{1}}{dx} y.  
\end{equation}
Using the volumetric flow rate condition, one can obtain the pressure gradient as 
\begin{equation}
P_{1}(x)\sim -\frac{\alpha^{2}+\eta^{2}}{\alpha^{2}} \frac{1}{H(x)}.    
\end{equation}
Similarly, one can obtain the $O(\delta^{2})$ velocity as 
\begin{equation}
 u_{p_{1}} \sim  \frac{\alpha^{2} (\alpha^{2} \lambda^{2}+ \eta^{2})}{2 (\alpha^{2}+\eta^{2})^{2}} \frac{d^{2}P_{1}}{dx^{2}} y^{2} +d_{1}(x),
\end{equation}
\begin{equation}
 d_{1}(x)\sim -\frac{\alpha^{2}(\alpha^{2}\lambda^{2}+\eta^{2})}{6 (\alpha^{2}+\eta^{2})^{2}} \frac{d^{2}P_{1}}{dx^{2}} H(x)^{2}.   
\end{equation}
Correspondingly, the axial velocity up to $O(\delta^{2})$ is given by 
\begin{equation}\label{Darcy velocity}
 u_{p}(x,y,t)\sim \left(u_{p_{0}}(x,y)+\delta^{2} u_{p_{1}}(x,y)\right) e^{i \textrm{Wo} t},  
\end{equation}
and the corresponding pressure gradient can be obtained from Eq. (\ref{x component pulsatile eqn}) using the Darcy velocity given in Eq. (\ref{Darcy velocity}).

\subsubsection{Flow reversal zone}
Based on the calculation presented in Sec. \ref{Darcy behavior far from wall}, one can get the approximate steady state Darcy velocity far from the wavy wall as 
\begin{equation}
u_{s}(x,y) \sim \frac{1}{H} + \frac{\lambda^{2} \delta^{2}}{6 H^3}  \left(2 \left(\frac{dH}{dx}\right)^{2}-\frac{d^{2}H}{dx^{2}}H\right) \left(H^{2}-3y^{2}\right), 
\end{equation}
and the corresponding pulsatile velocity as 
\begin{eqnarray}
u_{p}(x,y,t) \sim  \frac{\cos(\textrm{Wo} t)}{H}+ \frac{1}{6 (\alpha^{4}+\textrm{Wo}^{2}) H^{3}} \left(\delta^{2} \Sigma(t) \left(2 \left(\frac{dH}{dx}\right)^{2}-\frac{d^{2}H}{dx^{2}}H\right) \left(H^{2}-3y^{2}\right)\right)    
\end{eqnarray}
where,
\begin{equation}
 \Sigma(t)=  \left(\cos \left(\textrm{Wo}  t \right) \alpha^{4} \lambda^{2}+\sin \left(\textrm{Wo}  t \right) \alpha^{2} \lambda^{2} \textrm{Wo} -\sin \left(\textrm{Wo}  t \right) \alpha^{2} \textrm{Wo} +\cos \left(\textrm{Wo}  t \right) \textrm{Wo}^{2} \right).
\end{equation}
Note that $$\Sigma(t)\leq \left(\alpha^{2}+\textrm{Wo}\right) \left(\alpha^{2}\lambda^{2}+\textrm{Wo}\right).$$ Futhermore, under the assumption that $\textrm{Wo}\sim \alpha^{2}$, the above estimate is valid in the regime where $\lambda^{2}\delta^{2}\ll 1$. Accordingly, we have chosen the parameter $\lambda$ and $\delta$, such that $ \lambda^{2} \delta^{2} \ll 1$. 

When the flow is driven by the combined effects of steady and pulsating components, the resulting composite Darcy velocity far from the wavy wall is given by
\begin{equation}
 u(x,y,t) \sim u_{s}(x,y)+ \chi u_{p}(x,y,t).   
\end{equation}

\begin{table}[h!]
 \begin{center}
\begin{tabular}{   p{2.0cm} p{2.0cm} p{2.0cm} p{2.0cm}  p{5.5cm}  p{2.0cm}}
\hline
$t$ & $\lambda$ & $\delta^{2}$ & $\lambda^{2} \delta^{2}$ & $R_{\textrm{zone}}$ (Approx. flow reversal zone) &$H(0.25)$\\
\hline
$0$ & $1$ & $0.09$   & $0.09$ & $|y|>1.73$ (No circulation) & $1.3$\\
$0$ & $1.25$ & $0.09$   & $0.14$ & $|y|>1.50$ (No circulation) & $1.3$\\
$0$ & $1.5$ & $0.09$   & $0.20$ & $|y|>1.35$ (No circulation) & $1.3$\\
$0$ & $1.75$ & $0.09$   & $0.27$ & $|y|>1.23$ (Circulation) & $1.3$\\
$0$ & $2$ & $0.09$   & $0.36$ & $|y|>1.14$ (Circulation) & $1.3$\\
\hline
\hline
$T_{p}/4$ & $1$ & $0.09$   & $0.09$ & $|y|>1.73$ (No circulation) & $1.3$\\
$T_{p}/4$ & $1.25$ & $0.09$   & $0.14$ & $|y|>1.45$ (No circulation) & $1.3$\\
$T_{p}/4$ & $1.5$ & $0.09$   & $0.20$ & $|y|>1.28$ (Circulation) & $1.3$\\
$T_{p}/4$ & $1.75$ & $0.09$   & $0.27$ & $|y|>1.17$ (Circulation) & $1.3$\\
$T_{p}/4$ & $2$ & $0.09$   & $0.36$ & $|y|>1.08$ (Circulation) & $1.3$\\
\hline
\hline
$2T_{p}/4$ & $1$ & $0.09$   & $0.09$ & $|y|>1.73$ (No circulation) & $1.3$\\
$2T_{p}/4$ & $1.25$ & $0.09$   & $0.14$ & $|y|>1.37$ (No circulation) & $1.3$\\
$2T_{p}/4$ & $1.5$ & $0.09$   & $0.20$ & $|y|>1.18$ (Circulation) & $1.3$\\
$2T_{p}/4$ & $1.75$ & $0.09$   & $0.27$ & $|y|>1.07$ (Circulation) & $1.3$\\
$2T_{p}/4$ & $2$ & $0.09$   & $0.36$ & $|y|>1.0$ (Circulation) & $1.3$\\
\hline
\hline
$3T_{p}/4$ & $1$ & $0.09$   & $0.09$ & $|y|>1.73$ (No circulation) & $1.3$\\
$3T_{p}/4$ & $1.25$ & $0.09$   & $0.14$ & $|y|>1.18$ (Circualtion) & $1.3$\\
$3T_{p}/4$ & $1.5$ & $0.09$   & $0.20$ & $|y|>1.01$ (Circulation) & $1.3$\\
$3T_{p}/4$ & $1.75$ & $0.09$   & $0.27$ & $|y|>0.93$ (Circulation) & $1.3$\\
$3T_{p}/4$ & $2$ & $0.09$   & $0.36$ & $|y|>0.88$ (Circulation) & $1.3$\\
\hline
\hline
$4T_{p}/4$ & $1$ & $0.09$   & $0.09$ & $|y|>0.75$ (Circulation) & $1.3$\\
$4T_{p}/4$ & $1.25$ & $0.09$   & $0.14$ & $|y|>0.75$ (Circulation) & $1.3$\\
$4T_{p}/4$ & $1.5$ & $0.09$   & $0.20$ & $|y|>0.75$ (Circulation) & $1.3$\\
$4T_{p}/4$ & $1.75$ & $0.09$   & $0.27$ & $|y|>0.75$ (Circulation) & $1.3$\\
$4T_{p}/4$ & $2$ & $0.09$   & $0.36$ & $|y|>0.75$ (Circulation) & $1.3$\\
\hline
\hline
\end{tabular}
\end{center}   
 \end{table}  

\begin{table}[h!]
 \begin{center}
\begin{tabular}{  p{2.0cm} p{2.0cm} p{2.0cm} p{2.0cm}  p{5.5cm}  p{2.0cm}}
\hline
$t$ & $\lambda$ & $\delta^{2}$ & $\lambda^{2} \delta^{2}$ & $R_{\textrm{zone}}$ (Approx. flow reversal zone) &$H(0.25)$\\
\hline
\hline
$5T_{p}/4$ & $1$ & $0.09$   & $0.09$ & $|y|>1.73$ (No circulation) & $1.3$\\
$5T_{p}/4$ & $1.25$ & $0.09$   & $0.14$ & $|y|>1.45$ (No circulation) & $1.3$\\
$5T_{p}/4$ & $1.5$ & $0.09$   & $0.20$ & $|y|>1.28$ (No circulation) & $1.3$\\
$5T_{p}/4$ & $1.75$ & $0.09$   & $0.27$ & $|y|>1.2$ (can't predict accurately )  & $1.3$\\
$5T_{p}/4$ & $2$ & $0.09$   & $0.36$ & $|y|>1.09$ (can't predict accurately) & $1.3$\\
\hline
\hline
$6T_{p}/4$ & $1$ & $0.09$   & $0.09$ & $|y|>1.73$ (No circulation) & $1.3$\\
$6T_{p}/4$ & $1.25$ & $0.09$   & $0.14$ & $|y|>1.57$ (No circulation) & $1.3$\\
$6T_{p}/4$ & $1.5$ & $0.09$   & $0.20$ & $|y|>1.43$ (No circulation) & $1.3$\\
$6T_{p}/4$ & $1.75$ & $0.09$   & $0.27$ & $|y|>1.32$ (No circulation) & $1.3$\\
$6T_{p}/4$ & $2$ & $0.09$   & $0.36$ & $|y|>1.24$ (Onset of circulation ) & $1.3$\\
\hline
\hline
$7T_{p}/4$ & $1$ & $0.09$   & $0.09$ & $|y|>1.73$ (No circulation) & $1.3$\\
$7T_{p}/4$ & $1.25$ & $0.09$   & $0.14$ & $|y|>1.54$ (No circulation) & $1.3$\\
$7T_{p}/4$ & $1.5$ & $0.09$   & $0.20$ & $|y|>1.40$ (No circulation) & $1.3$\\
$7T_{p}/4$ & $1.75$ & $0.09$   & $0.27$ & $|y|>1.3$ (No circulation) & $1.3$\\
$7T_{p}/4$ & $2$ & $0.09$   & $0.36$ & $|y|>1.20$ (Circulation) & $1.3$\\
\hline
\hline
\end{tabular}
\end{center}   
\caption{Approximate flow reversal zone beneath the crest ($x = 0.25$) of the wavy wall for the full period of pulsation when the flow is driven by a combination of steady and pulsatile components, for the parameter values $\delta=0.3$, $a = 0.3$, $Da = 0.01$, $\chi=1$, and $\textrm{Wo} = 100$.}\label{Table_data}
\end{table} 

Motivated by the observation that a circulation zone occurs at the crest region beneath the hollow part of the wavy wall, which is particularly critical for mass transport and enhanced solute mixing, we have estimated the approximate flow reversal zone where circulations occur, as presented in Table \ref{Table_data}.
Based on the calculations presented in Sec. \ref{Darcy behavior far from wall}, we determined the flow reversal zone at various time instances throughout a complete pulsation period for different anisotropic ratios $\lambda$. According to Darcy's approximation theory, the estimated flow reversal zone yields qualitatively good results, as shown in Figs. (\ref{Velocity t=0}-\ref{Velocity t=7T/4}).

\begin{figure}[htp!]
    \centering
    \subfigure[Axial velocity at $x=0.25$]{\label{Darcy Brinkman velocity}\includegraphics[height=2.6 in, width=2.8 in]{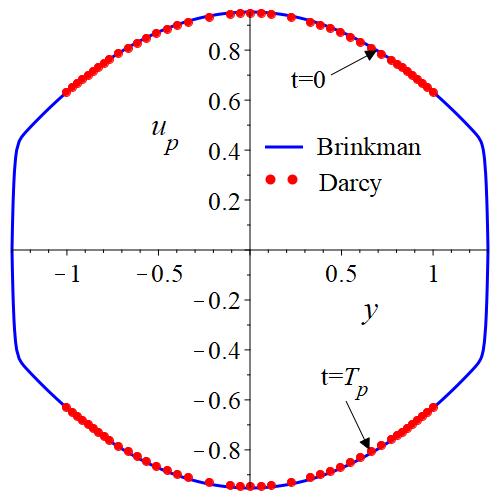}}
    \subfigure[Pressure gradient at the wall]{\label{Darcy Brinkman Pressure gradient}\includegraphics[height=2.6 in, width=2.8 in]{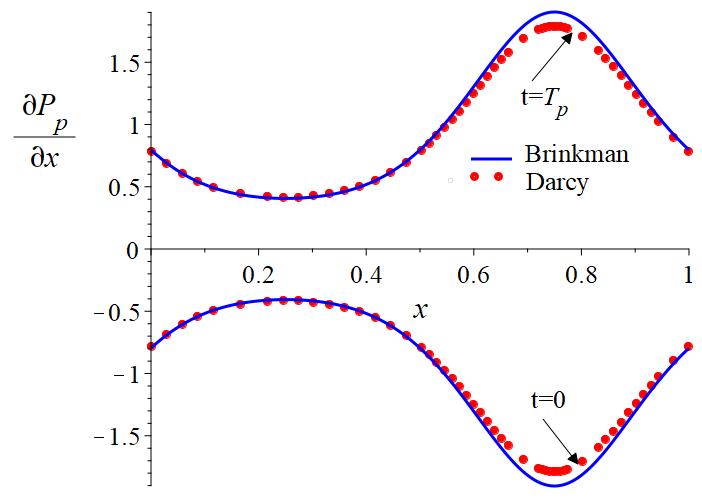}}
    \caption{Axial velocity and pressure gradient profile for the parameter values $\delta=0.3, a=0.3, Da=0.001, M=1, \lambda=1, \textrm{Wo}=100$.}
    \label{Darcy Brinkman velocity and pressure gradient}
\end{figure}

As discussed previously, the flow satisfies the Darcy equation far from the wavy wall, while it follows the Brinkman equation in the region near the wall. A similar behavior can be expected in flows through a low-permeable porous medium. To explore the interplay between Brinkman and Darcy effects, we have plotted the axial velocity distribution in Fig. (\ref{Darcy Brinkman velocity}) and pressure gradient in Fig. (\ref{Darcy Brinkman Pressure gradient}). In both cases, the Brinkman solution near the wavy wall closely matches the Darcy behavior far from the wavy wall. This indicates the presence of a thin layer (say, the Brinkman layer) next to the wavy boundary, where viscous effects are evident, and the flow field follows the Brinkman equation.



\section{Some special cases} In case of unperturbed flow, $a \sim 0$, implying $H(x)=1.$ In this we get the axial velocity for the case of pulsatile flow as 
\begin{equation}
 u_{p_{0}}(y,t)=\frac{\left(\cosh(y\sqrt{\alpha^{2}+\eta^{2}} )-\cosh(\sqrt{\alpha^{2}+\eta^{2}} )\right)\sqrt{\alpha^{2}+\eta^{2}} }{\left(\sinh(\sqrt{\alpha^{2}+\eta^{2}})-\sqrt{\alpha^{2}+\eta^{2}} \cosh(\sqrt{\alpha^{2}+\eta^{2}})\right)} e^{i \textrm{Wo} t}.   
\end{equation}

In case of unperturbed flow, i.e., for $\delta=0, a=0$, if the permeability in the horizontal direction is very large, then $\alpha \ll 1$, and, 
\begin{equation}
 \cosh\left(\sqrt{\alpha^{2}+\eta^{2}}\right)\sim \frac{\left(e^{\csgn(\eta) \eta}\right)^{2}+1}{2 e^{\csgn(\eta) \eta}} +O(\alpha^{2}),
\end{equation}
\begin{equation}
  \sinh\left(\sqrt{\alpha^{2}+\eta^{2}}\right)   \sim \frac{\left(e^{\csgn(\eta) \eta}\right)^{2}-1}{2 e^{\csgn(\eta) \eta}} +O(\alpha^{2}),
\end{equation}
\begin{equation}
 \sqrt{\alpha^{2}+\eta^{2}}\cosh\left(\sqrt{\alpha^{2}+\eta^{2}}\right)\sim \frac{\left(\left(e^{\csgn(\eta) \eta}\right)^{2}+1\right)\csgn(\eta) \eta}{2 e^{\csgn(\eta) \eta}} +O(\alpha^{2}).   
\end{equation}
Consequently, we get 
\begin{equation}
 u_{p_{0}}(y,t)\sim \frac{\left(\frac{\left(e^{\csgn(\eta) \eta y}\right)^{2}+1}{2 e^{\csgn(\eta) \eta y}}-\frac{\left(e^{\csgn(\eta) \eta}\right)^{2}+1}{2 e^{\csgn(\eta) \eta}}\right) \csgn(\eta) \eta}{\left(\frac{\left(e^{\csgn(\eta) \eta}\right)^{2}-1}{2 e^{\csgn(\eta) \eta}}-\frac{\left(\left(e^{\csgn(\eta) \eta}\right)^{2}+1\right)\csgn(\eta) \eta}{2 e^{\csgn(\eta) \eta}}\right)} e^{i \textrm{Wo} t}+O(\alpha^{2}),   
\end{equation}
which is nothing but the velocity profile for pulsatile flow in the case of clear flow situation \cite{Loudon1998use}, where the $\csgn(x)$ is defined by 
\begin{equation}
    \csgn(x) = \begin{cases}
        1, & 0 <\Re e (x)~ \textrm{or} ~\Re e (x)=0~ \textrm{and}~ 0<\Im(x),\\
        -1, & \Re e (x)<0 ~ \textrm{or} ~ \Re  e (x)=0~ \textrm{and}~\Im(x)<0.
    \end{cases}
\end{equation}

\begin{figure}[h!]
    \centering
    \subfigure[]{\label{Time avegarage pressure gradient}\includegraphics[height=2.5 in, width=3.0 in]{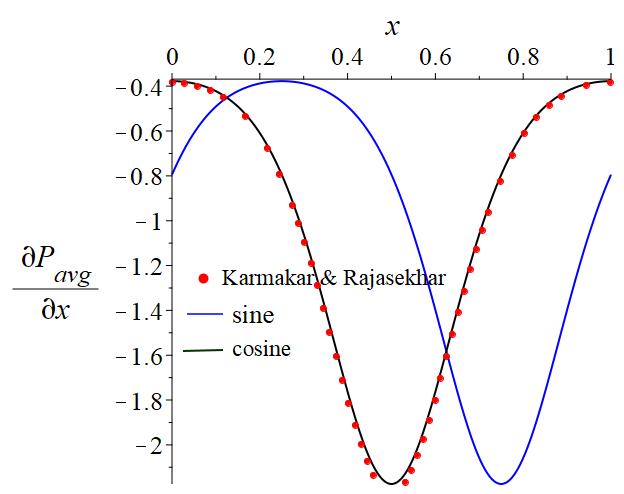}}
    \subfigure[]{\label{Time average shear stress}\includegraphics[height=2.5 in, width=3.0 in]{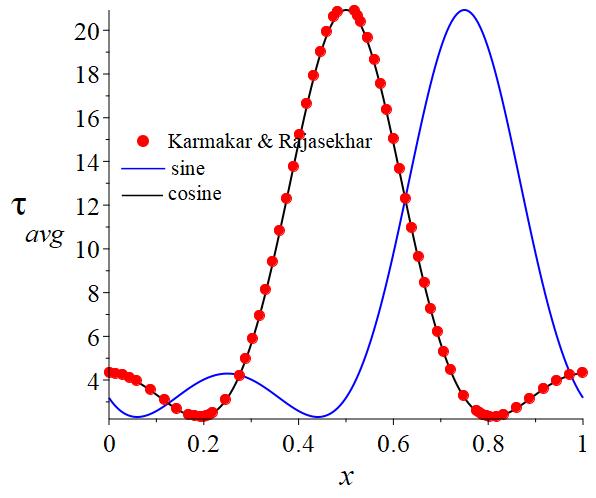}}
    \caption{Time-averaged (a) pressure gradient, (b) shear stress distribution at the wall for the parameters $\delta=0.3, a=0.3, Da=0.01, M=1$, $ \lambda=1, \textrm{Wo}=1$ and $\chi=1$.}
    \label{Time average pressure and shear stress}
\end{figure}

A comparison of the present study with previous work is depicted in Fig. \ref{Time average pressure and shear stress}. The time-averaged pressure gradient and shear stress are shown when the wavy wall corresponds to both sinusoidal and cosinusoidal curves (refer to Figs. \ref{Time avegarage pressure gradient}-\ref{Time average shear stress}). Our results agree with the findings of Karmakar and Rajasekhar \cite{Karmakar2017note} when the wavy wall follows a cosinusoidal profile.

\section{Wall shear stress distribution for the case of large Darcy number: application to blood flow in arteries}

\begin{figure}[h!]
    \centering
    \subfigure[$t=0$]{\label{Dialation shear stress}\includegraphics[height=2.2 in, width=2.7 in]{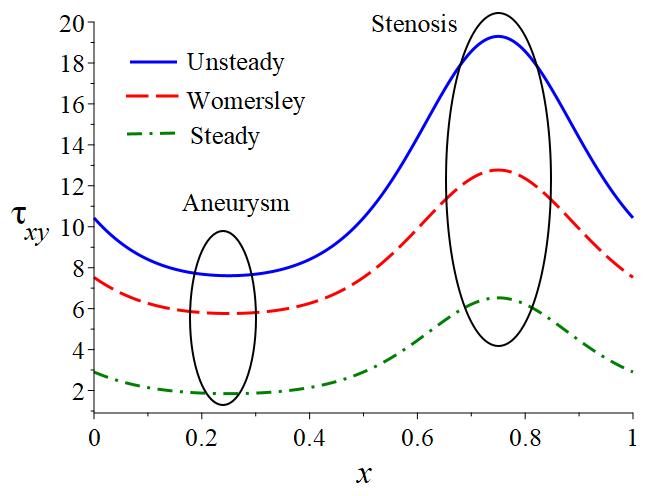}}
    \subfigure[$t=T_{p}$]{\label{Stenosis shear stress}\includegraphics[height=2.2 in, width=2.7 in]{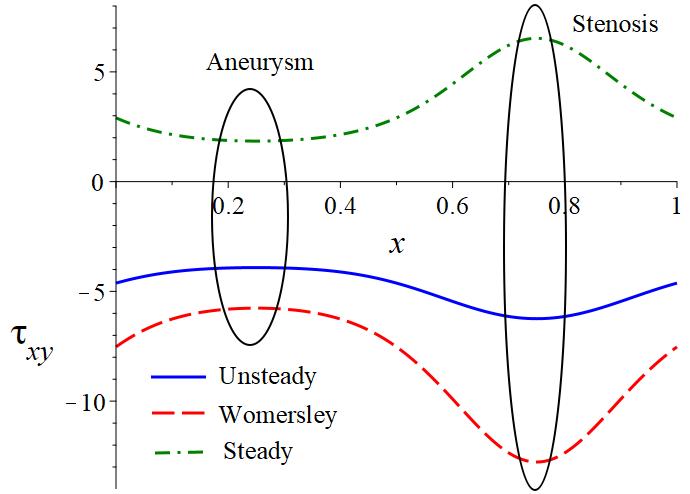}}
    \caption{Shear stress distribution as a function of $x$ for $Da=1$, $M=1$, $\lambda=1$, $a=0.3$, $\delta=0.1$,  $\textrm{Wo}=100$.}
    \label{Dialation and Stenosis shear stress}
\end{figure}

\begin{figure}[h!]
    \centering
    \subfigure[$x=0.25,t=0$]{\label{Velocity three case 1}\includegraphics[height=2.2 in, width=2.2 in]{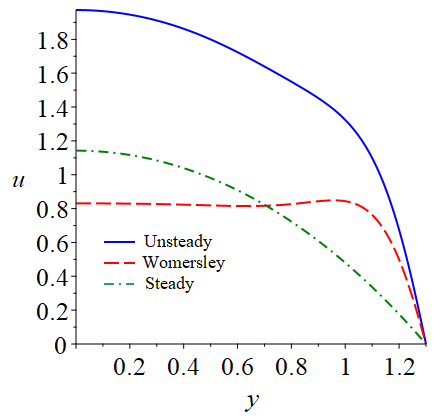}}
    \subfigure[$x=0.75,t=0$]{\label{Velocity three case 2}\includegraphics[height=2.2 in, width=2.2 in]{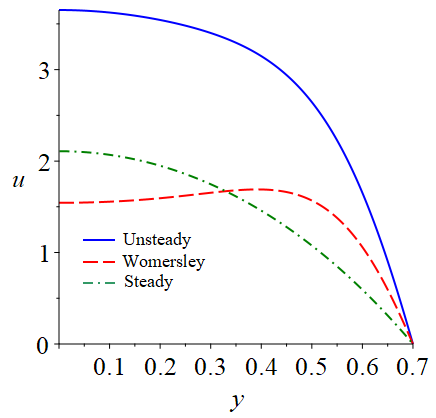}}
    \subfigure[$x=0.25,t=T_{p}$]{\label{Velocity three case 3}\includegraphics[height=2.0 in, width=2.2 in]{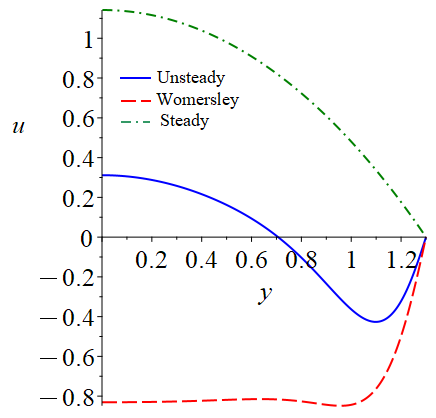}}
    \subfigure[$x=0.75,t=T_{p}$]{\label{Velocity three case 4}\includegraphics[height=2.0 in, width=2.2 in]{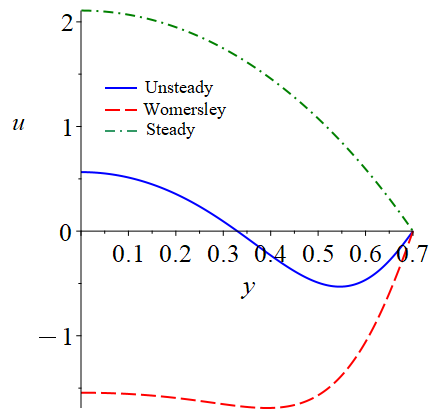}}
    \caption{Velocity profile at the aneurysmal and stenotic regions for the parameter values $Da=1$, $M=1$, $\lambda=1$, $a=0.3$, $\delta=0.1$, $\textrm{Wo}=100$.}
    \label{Velocity three case}
\end{figure}

Pulsatile blood flow through a wavy-walled channel has drawn the interest of researchers as it serves as a simplified model to mimic the hemodynamics observed in stenotic (narrowed) and aneurysmal (dilated) blood vessels. Studying pulsatile flow in a wavy channel provides valuable insight into understanding how geometric irregularities in arteries affect the wall shear stress (WSS) distribution, which may contribute to the development of cardiovascular disease, such as atherosclerosis. Blood flow in a constricted artery can be compared to fluid flow through a wavy channel (clear flow situation), where the narrow troughs represent areas of arterial blockage \cite{Mittal2003numerical,Bandyopadhyay2012study}. This simplified model helps us to understand how stenosis affects hemodynamics. A reduction in arterial cross-sectional area of more than $70\%$ is typically considered to be clinically significant as it severely impairs blood flow and poses a serious risk to the patient \cite{Mittal2003numerical,Carretta1998mcdonald}. In view of this, we have plotted the wall shear stress distribution as a function of position $x$ in Fig. \ref{Dialation and Stenosis shear stress} corresponding to the locations of geometric irregularities along the channel. Since a wavy channel with clear flow conditions is more relevant in the current context, we have analyzed the corresponding wall shear stress distribution for the case of a large Darcy number (i.e., $Da=1$), where the flow closely resembles that in a non-porous medium. Figure \ref{Dialation and Stenosis shear stress} illustrates the relationship between the shear stress distribution and variations in the geometry of the channel, particularly in the presence of stenosis and aneurysms. In stenotic regions, the narrowing of blood vessels accelerates blood flow, resulting in a higher velocity magnitude (refer to Figs. \ref{Velocity three case 1}-\ref{Velocity three case 4}) and increased velocity gradients. This leads to significantly elevated WSS, a phenomenon that has been observed in previously attempted studies \cite{Dawood2024effect,Dawood2024pulsatile,Abdelsalam2020alterations}. In contrast, in aneurysmal regions where blood vessels expand, flow retardation and the formation of recirculation zones result in reduced shear stress. Therefore, the dynamic interplay between geometry-induced flow alterations and variations in WSS is a key factor in understanding the vascular pathophysiology. In Fig. \ref{Dialation shear stress}, we have plotted the shear stress distribution at the time $t=0$, while in Fig. \ref{Stenosis shear stress}, it is plotted for time $t=T_{p}$. At time $t=0$, the shear stress remains positive across the entire length of the artery. The maximum shear stress occurs at the stenotic region (i.e., at $x=0.75$), while the minimum is found at the aneurysmal region (i.e., at $x=0.25$). In this scenario, the unsteady shear stress is significantly higher compared to both the Womersley and steady cases. This occurs because of the positive velocity distribution for the pulse $t=0$ (see Figs. \ref{Velocity three case 1}-\ref{Velocity three case 2}). However, for the pulse at $t=T_{p}$, although the steady state shear stress distribution stays positive, both unsteady and Womersley cases show negative shear stress values across the entire arterial length. This change is attributed to the alterations in velocity and backflow characteristics of the unsteady and Womersley condition (see Figs. \ref{Velocity three case 3}-\ref{Velocity three case 4}).


\section{Conclusions}

We studied the combined effect of steady and pulsatile flow within the wavy channel filled with an anisotropic porous medium. The hydrodynamic phenomenon is examined by solving the governing equation with the perturbation method. We studied the influence of various factors on both steady and pulsatile flow by presenting three distinct cases: the steady case, the pulsatile case, and the combined effect of steady and pulsatile flow. We have shown a significant dependence of the anisotropic permeability ratio ($\lambda$) and the time ($t$) on flow velocity, streamline, pressure gradient, and shear stress. Our analysis indicates that in the steady state scenario, the eddy structure forms near the crest of the wavy channel as the anisotropic permeability ratio increases. However, when examining the pulsatile effect, flow circulation can be seen for an isotropic situation as well for certain time instances. For higher values of anisotropic permeability ratio, $\lambda$, vortices have been observed to form not only at the crest but also near the trough region of the wavy wall. The study indicates that the pulsatile effect significantly influences flow separation in both isotropic and anisotropic porous media. For the case of a large Darcy number, we analyzed the shear stress distribution along the wavy wall to replicate flow conditions in stenotic and aneurysmal arteries. The results show that shear stress is significantly elevated in the stenotic region and reduced in the aneurysmal region. This insight contributes to a better understanding of how arterial wall damage occurs and blockages develop. The flow separation phenomenon has a significant use in membrane oxygenators for cardiopulmonary bypass, where a porous packing is utilized to separate the blood from the oxygen and minimize blood damage from direct oxygen contact. The present study demonstrates that the anisotropic permeability of the porous material and the pulsation period contribute to the flow separation in both the crest and trough regions of the wavy channel. These findings may facilitate the separation of oxygen and blood and improve the efficacy of oxygenators.
\vspace{0.2 cm}

\noindent \textbf{Acknowledgments:} The first author acknowledges the institute research fellowship of National Institute of Technology Meghalaya.


\providecommand{\noopsort}[1]{}\providecommand{\singleletter}[1]{#1}%

\end{document}